\documentclass[sigconf]{acmart}
\AtBeginDocument{%
  }

\copyrightyear{2026}
\acmYear{2026}
\setcopyright{none}
\acmConference[WWW '26]{Proceedings of the ACM Web Conference 2026}{April 13--17, 2026}{Dubai, United Arab Emirates}
\acmBooktitle{Proceedings of the ACM Web Conference 2026 (WWW '26), April 13--17, 2026, Dubai, United Arab Emirates}
\acmDOI{10.1145/3774904.3792180}
\acmISBN{979-8-4007-2307-0/2026/04}

\usepackage{soul}
\usepackage{graphicx}
\usepackage{amsmath}
\usepackage{amsthm}
\usepackage{algorithm}
\usepackage{algorithmic}
 
\usepackage{caption} 
\usepackage{multirow} 
\usepackage{subcaption}  
\usepackage{enumitem}

\begin{document}

\title{negMIX: Negative Mixup for OOD Generalization in Open-Set Node Classification}


\author{Junwei Gong}
\orcid{0009-0000-0218-8105}
\affiliation{%
  \institution{Hainan University}
  \city{Haikou}
  \state{Hainan}
  \country{China}}
\email{gjw@hainanu.edu.cn}

\author{Xiao Shen}
\orcid{0000-0003-0937-049X}
\authornote{Corresponding Author.}
\affiliation{%
  \institution{Hainan University}
  \city{Haikou}
  \state{Hainan}
  \country{China}}
\email{xshen@hainanu.edu.cn}

\author{Zhihao Chen}
\orcid{0000-0002-0107-0663}
\affiliation{%
  \institution{Hainan University}
  \city{Haikou}
  \state{Hainan}
  \country{China}}
\email{zhchen@hainanu.edu.cn}

\author{Shirui Pan}
\orcid{0000-0003-0794-527X}
\affiliation{%
  \institution{Griffith University}
  \city{Queensland}
  \state{Gold Coast}
  \country{Australia}}
\email{s.pan@griffith.edu.au}

\author{Xiao Wang}
\orcid{0000-0002-4444-7811}
\affiliation{%
  \institution{Beihang University}
  \city{Beijing}
  \country{China}}
\email{xiao\_wang@buaa.edu.cn}

\author{Xi Zhou}
\orcid{0000-0002-6943-5585}
\affiliation{%
  \institution{Hainan University}
  \city{Haikou}
  \state{Hainan}
  \country{China}}
\email{xzhou@hainanu.edu.cn}

\renewcommand{\shortauthors}{Junwei Gong et al.}

\begin{abstract}
Open-set node classification (OSNC) allows unlabeled test data to contain novel classes previously unseen in the labeled data. The goal is to classify in-distribution (ID) nodes into corresponding known classes and reject out-of-distribution (OOD) nodes as unknown class. Despite recent notable progress in OSNC, two challenges remain less explored, i.e., how to enhance generalization to OOD nodes, and promote intra-class compactness and inter-class separability. To tackle such challenges, we propose a novel Negative Mixup with Cross-Layer Graph Contrastive Learning (negMIX) model. Firstly, we devise a novel negative Mixup method purposefully crafted for the open-set scenario with theoretical justification, to enhance the model's generalization to OOD nodes and yield clearer ID/OOD boundary. Additionally, a unique cross-layer graph contrastive learning module is developed to maximize the prototypical mutual information between the same class nodes across different topological distance neighborhoods, thereby facilitating intra-class compactness and inter-class separability. Extensive experiments validate significant outperformance of the proposed negMIX over state-of-the-art methods in various scenarios and settings.
\end{abstract}



\maketitle

\section{Introduction}
Graph neural networks (GNNs) have demonstrated promising performance on semi-supervised node classification (SSNC) task across \textbf{Web-related graphs}. Existing GNNs for SSNC are typically under a closed-set assumption \cite{GAT,DSA-GNN,CDNE,ACDNE,DM-GNN,AdaGCN}, where unlabeled test data contain only classes previously encountered in the labeled training data. Such a restrictive assumption significantly limits the applicability of these methods in dynamic graph environments, where out-of-distribution (OOD) nodes from novel classes may frequently arise at test time. Open-Set Node Classification (OSNC) \cite{OpenWGL}, which aims to accurately classify in-distribution (ID) nodes into corresponding known classes and reject OOD nodes as ``unknown'' class, has gained increasing attention in recent few years, due to its wide application in real-world scenarios. Despite recent notable progress in OSNC \cite{topoOOD,LMN,MHGL}, two critical challenges remain less explored.

\begin{figure}
\centering
\includegraphics[width=\linewidth]{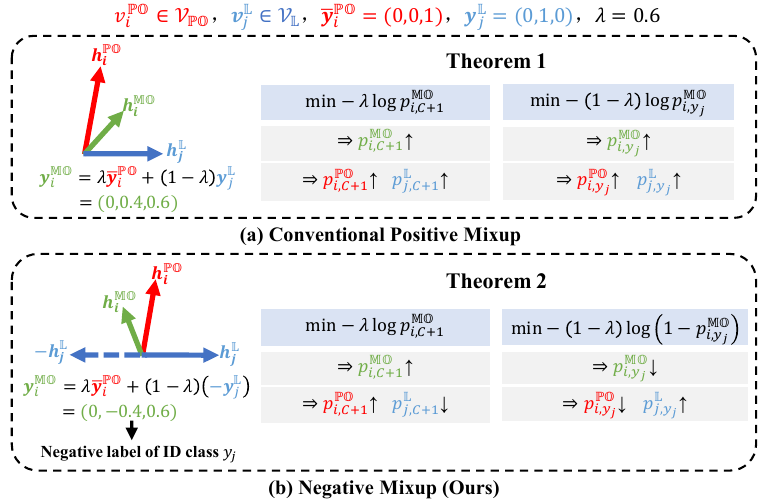}
\caption{Comparison between Conventional Positive Mixup and the proposed Negative Mixup. (a) Conventional Positive Mixup constructs a mixed-up OOD sample that lies between a labeled ID node and a potential OOD node, and assigns it with a soft positive label of both ID and OOD classes. (b) Our Negative Mixup constructs a mixed-up OOD sample close to OOD while far away from ID, and assigns it with a positive label of OOD class and a negative label of ID class.}
\Description{The figure contrasts conventional positive Mixup and the proposed Negative Mixup in the feature space. Conventional Mixup generates mixed samples between labeled ID nodes and potential OOD nodes and assigns them soft positive labels for both ID and OOD classes, which couples ID and OOD samples during optimization by simultaneously increasing known and unknown class probabilities, resulting in ambiguous ID/OOD decisions. In contrast, Negative Mixup constructs mixed samples that are closer to OOD embeddings while remaining far from ID embeddings, and assigns them a positive label for the OOD class and a negative label for the ID class. This induces mutually exclusive decision behavior by enhancing unknown class probability while suppressing known class probability for potential OOD samples, and vice versa for labeled ID samples, leading to a more discriminative decision boundary.}
\label{fig.negmixup}
\end{figure}

(1) \textbf{{Lack of OOD Supervised Signals}}. In OSNC setting, the supervised signals are only available from ID nodes during training, making the model exhibit poor generalization to OOD nodes \cite{OODdetect,AugAN}. Recently, Mixup techniques \cite{mixup,cutmix,tokenmix}, which combine different training samples and their labels to smooth the representations of original training data, have proven effective in enhancing model’s generalization and reducing over-fitting. However, standard Mixup approaches are mostly designed for the closed-set scenario that only contains ID samples, and they typically mix two ID samples \textbf{positively} to construct a new training sample. We argue that \textbf{conventional positive Mixup paradigm would be inappropriate for OOD generalization in OSNC problem}, where a graph contains both ID and OOD nodes. As illustrated in Figure \ref{fig.negmixup}a, if one adopts conventional Mixup to positively combine an original ID node with a potential OOD node to construct a new mixed-up pseudo-OOD training sample, then such constructed pseudo-OOD training sample would lie between original ID and OOD nodes. It is worth noting that the construction of clear decision boundaries depends on high quality and diversity of training samples. Thus, utilizing such mixed-up pseudo-OOD sample which has been smoothed between ID and OOD nodes for model training would unavoidably blur the ID and OOD boundary. Recent study \cite{openmix} also reveals that conventional Mixup would increase the structural risk and open space risk in the open-set scenario, thereby decaying both ID classification and OOD detection performance. However, so far, relatively less effort has been devoted to developing Mixup technique suitable for the open-set scenario. This presents the key motivation of this work: 
\textbf{How to develop an effective Mixup method explicitly tailored for the open-set scenario, to construct high-quality and diverse pseudo-OOD training samples for OSNC?} 

(2) \textbf{Intra-class Variety and Inter-class Confusion}. Due to the intrinsic connectivity of graph-structured data, possible message-passing between ID and OOD nodes in the OSNC problem would exacerbate the intra-class variety and inter-class confusion issue \cite{OODGAT,LMN,jiang2026scalable}, i.e., nodes within the same class may display diverse and scattered representations, while nodes from different classes may exhibit overlapping representations, resulting in ambiguous decision boundaries. Graph contrastive learning (GCL) \cite{nodedrop,edgeper} has demonstrated great success in learning informative embeddings for closed-set node classification task. However, conventional GCL methods may suffer from two limitations when directly applied to the OSNC task. On one hand, they typically generate contrastive views by handcraft augmentations, which may degrade the quality of embeddings \cite{NCLA}. On the other hand, they typically consider the same node of the anchor across views as positives while different nodes as negatives, failing to guarantee the class-discriminative requirement of OSNC. Hence, there exists \textbf{a compelling need to develop an effective GCL method to enhance intra-class compactness and inter-class separability for OSNC.}

To address the above challenges, we propose a novel Negative Mixup with Cross-Layer Graph Contrastive Learning (negMIX) model. Firstly, unlike conventional Mixup that positively mixes two samples, we propose a novel negative Mixup paradigm specifically crafted for the open-set scenario, which constructs a new pseudo-OOD training sample by mixing a selective potential OOD node up with the inverse vector of a labeled ID node at both embedding- and label-levels, as in Figure \ref{fig.negmixup}b. Moreover, we carefully devise the positive learning loss w.r.t. unknown class and negative learning loss w.r.t. known class on such negatively mixed-up pseudo-OOD samples for model training. As a result, the model's generalization to OOD nodes can be enhanced, and clearer boundary between ID and OOD can be yielded. In light of the second challenge, we propose a unique cross-layer GCL module, which considers different GNN layers inherently capturing various multi-hop information as learnable graph augmented views, and defines positives and negatives upon the class information. As a result, the shared information between the same class nodes across different topological distance neighborhoods can be maximized, so as to promote intra-class compactness and inter-class separability for both ID and OOD classes, facilitating clearer ID/OOD boundary for OSNC. Our contributions are summarized as:%

\begin{itemize}[leftmargin=*]
    \item We theoretically and practically demonstrate that conventional positive Mixup is infeasible for OOD generalization in the open-set scenario. To the best of our knowledge, we make the first attempt to develop a novel \textbf{negative Mixup} paradigm purposefully crafted for the OSNC problem with theoretical justification, which effectively enhances generalization to OOD nodes and facilitates clearer ID and OOD boundary.
    \item To enhance intra-class compactness and inter-class separability in OSNC, we introduce a unique \textbf{cross-layer GCL} module to maximize the prototypical mutual information across different distance neighborhoods, which defines positives and negatives upon class information across different GNN layers.
    \item Extensive experiments on benchmark graph datasets demonstrate the superior performance of the proposed negMIX over state-of-the-art (SOTA) OSNC methods under different training ratios and different numbers of OOD classes. Under the same condition, our negMIX improves the average Accuracy by 8.00\%, Macro-F1 by 6.77\%, AUROC by 5.98\% and substantially reduces the average FPR@95 by 18.69\% over the strongest baseline on eight datasets. 
\end{itemize}

\section{Related Work}
\textbf{Open-set Node Classification}. OpenWGL \cite{OpenWGL} leverages a variational graph auto-encoder to learn representations sensitive to unknown class. OODGAT \cite{OODGAT} designs an attention mechanism to model interactions between nodes from known and unknown classes. {G$^2$Pxy} \cite{G2Pxy} simulates unknown class distribution by generating pseudo-unknown class proxies. GOLD \cite{GOLD} proposes an implicit adversarial learning strategy to synthesize pseudo-OOD samples. GNNSAFE++ \cite{gnnsafe} proposes energy belief propagation to enlarge the energy gap between ID and OOD nodes. NODESAFE++ \cite{nodesafe} improves on GNNSAFE++ by bounding negative energy scores. GRASP \cite{grasp} designs a graph augmentation strategy to add intra-edges, improving OOD detection performance after OOD score propagation. In contrast to existing methods, our work provides a novel perspective for OSNC, with a negative Mixup method crafted for open-set scenario for effective OOD generalization, and a cross-layer GCL module to extract prototypical mutual information across different distance neighborhoods. 

\textbf{Graph Contrastive Learning} methods typically employ handcraft augmentations, such as node dropping \cite{nodedrop}, edge perturbation \cite{edgeper,GRASS}, attribute masking \cite{attrmask,GOUDA}, and graph diffusion \cite{graphdiffusion1} to generate augmented views with discrepancy. However, such handcraft augmentations may heavily damage the downstream task-relevant topology or attributes, thereby degrading the quality of embeddings \cite{NCLA,HEATS}. In addition, conventional GCL methods typically define positives as the same node across views while different nodes as negatives, failing to directly encourage intra-class compactness and inter-class separation. Compared to previous GCL methods, two key distinctions of our work lie in: 1) To get rid of handcraft graph augmentations, we consider different GNN layers inherently capturing various distance neighborhoods as learnable augmentation. 2) We define positive and negative pairs upon the class information, promoting intra-class compactness and inter-class separability.

\textbf{Mixup} combines different original training samples and their labels to encourage the model to learn more generalized embeddings, 
thereby reducing over-fitting and enhancing generalization \cite{mixup}. Existing Mixup techniques \cite{manifoldmix,cutmix,tokenmix} are mostly designed for the closed-set scenario. However, they have been proven ineffective in the open-set scenario, since they would smooth the score function between ID and OOD samples, blurring the decision boundaries, thereby increasing the structural risk and open space risk \cite{openmix}. Very few latest studies \cite{openmix,openmix1} propose the modified Mixup techniques for the open-set scenario. However, they primarily focus on combining ID samples at the embedding-level, while neither considering label mixing nor designing a principled loss to train such mixed-up samples. To overcome this, our work devises a pioneering negative Mixup method explicitly tailored for open-set scenario with theoretical justification, to enhance OOD generalization and facilitate clearer ID and OOD boundary.

\begin{figure}
\setlength{\abovecaptionskip}{6pt}
\setlength{\belowcaptionskip}{-6pt}
\centering
\includegraphics[width=\linewidth]{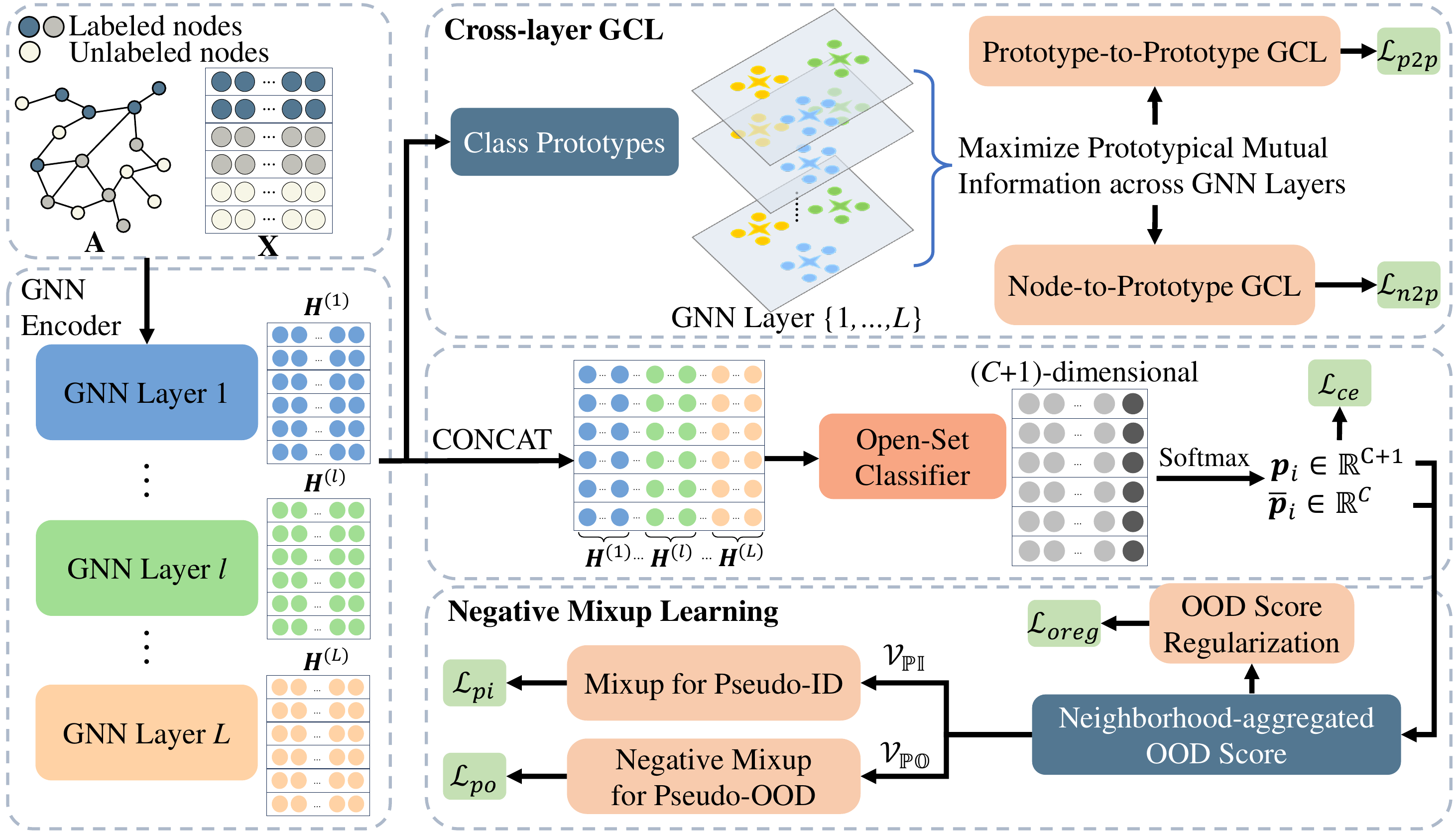}
\caption{Model architecture of negMIX which consists of negative Mixup and cross-layer GCL.}
\Description{The figure illustrates the overall architecture of negMIX. Negative Mixup Learning constructs new training samples by mixing labeled-ID nodes up with potential-ID (OOD) nodes selected upon neighborhood-aggregated OOD score. Cross-layer GCL maximizes prototypical mutual information across different GNN layers at prototype-to-prototype and node-to-prototype levels.}
\label{fig.model}
\end{figure}

\section{Preliminaries}
Let $G=(\mathcal{V},\mathcal{E},\boldsymbol{X},\boldsymbol{A},\boldsymbol{Y})$ be a graph, where $\mathcal{V}$ and $\mathcal{E}$ denote the node set and the edge set. $\boldsymbol{A}\in\mathbb{R}^{N\times N}$ is the adjacency matrix, where $N$ is the number of nodes, $A_{ij}=1$ if there is an edge between nodes $v_i$ and $v_j;$ otherwise, $A_{ij}=0$. $\boldsymbol{X}\in\mathbb{R}^{N\times F}$ and $\boldsymbol{Y}\in\mathbb{R}^N$ denote the feature matrix and the label vector, where $F$ is the number of features. Each node $v_i$ is associated with a feature vector $\boldsymbol{x}_i$ and a label $y_i.$ In SSNC, the node set $\mathcal{V}$ is divided into two distinct sets $\mathcal{V}=$ $\mathcal{V}_\mathrm{\mathbb{L}}\cup\mathcal{V}_\mathrm{\mathbb{U}}$, where $\mathcal{V}_\mathrm{\mathbb{L}}$ represents the labeled set with accessible labels during training, and $\mathcal{V}_\mathrm{\mathbb{U}}$ represents the unlabeled set. The label vector $\boldsymbol{Y}$ can be also divided as $\boldsymbol{Y}=[\boldsymbol{Y}_\mathrm{\mathbb{L}}\|\boldsymbol{Y}_\mathrm{\mathbb{U}}].$ Let $\mathcal{C}_\mathrm{\mathbb{L}}$ and $\mathcal{C}_\mathrm{\mathbb{U}}$ denote the set of classes associated with $\boldsymbol{Y}_{\mathbb{L}}$ and $\boldsymbol{Y}_{\mathbb{U}}$ respectively. Traditional closed-set SSNC problem assumes $\mathcal{C}_\mathrm{\mathbb{L}} = \mathcal{C}_\mathrm{\mathbb{U}}$.

\textbf{Open-set Node Classification} problem considers a more realistic scenario, where some unlabeled nodes in $\mathcal{V}_\mathrm{\mathbb{U}}$ may not belong to any classes seen in the labeled set, i.e., $\mathcal{C}_{\mathbb{L}}\neq \mathcal{C}_{\mathbb{U}}$ and $\mathcal{C}_\mathrm{\mathbb{L}}\subset \mathcal{C}_\mathrm{\mathbb{U}}.$ Let $\mathcal{C}_\mathrm{\mathbb{U}}\setminus \mathcal{C}_\mathrm{\mathbb{L}}$ denote a set of novel categories unseen in the labeled set, which may contain multiple classes. Nodes with labels belonging to $\mathcal{C}_\mathrm{\mathbb{L}}$ are regarded as ID nodes, and  belonging to $\mathcal{C}_\mathrm{\mathbb{U}}\setminus \mathcal{C}_\mathrm{\mathbb{L}}$ are regarded as OOD nodes. The goal of OSNC is to learn a ($C$+1)-class classifier $f(\mathcal{V}_{\mathbb{L}},\mathcal{V}_{\mathbb{U}},\boldsymbol{X},\boldsymbol{A},\boldsymbol{Y}_{\mathbb{L}})\mapsto$ $\{1,...,C,unknown\}$, where $C$ is the number of known classes in $\mathcal{C}_\mathrm{\mathbb{L}}$. Specifically, 1) for known class classification, each ID node in $\mathcal{V}_\mathrm{\mathbb{U}}$ should be correctly classified into one of pre-$C$ known classes; and 2) for unknown class detection, OOD nodes in $\mathcal{V}_\mathrm{\mathbb{U}}$ should be recognized as the ($C$+1)-th “unknown” class. Note that this work studies OSNC under the transductive learning setting \cite{OODGAT}, i.e., the whole graph (including unlabeled nodes) is accessible during model training.  

\section{The Proposed negMIX Model}
\label{sec:method}
The framework of negMIX is illustrated in Figure \ref{fig.model}, which consists of two key modules, i.e., 1) negative Mixup to enhance OOD generalization, and 2) cross-layer GCL to promote intra-class compactness and inter-class separability.

\textbf{Open-set Classifier.} Graph Attention Network (GAT) \cite{GAT} is adopted as the GNN encoder. While in contrast to conventional GAT, the proposed negMIX concatenates the embeddings at each GNN layer to obtain the final embedding to capture both local and global information as: $\boldsymbol{h}_i=\mathrm{CONCAT}(\boldsymbol{h}_i^{(1)},...,\boldsymbol{h}_i^{(L)})$, $\boldsymbol{h}_i^{(l)}=\sigma(\boldsymbol{h}_{i}^{(l-1)},\mathrm{AGG}(\boldsymbol{h}_{j}^{(l-1)};j\in\mathcal{N}_i))$, where $\boldsymbol{h}_i^{(l)}$ is the embedding of $v_i$ at the $l$-th GNN layer capturing the information from neighbors that are $l$-hop away, $L$ is the number of GNN layers, $\mathcal{N}_i$ is a set of first-order neighbors of $v_i$, $\mathrm{AGG}(\cdot)$ and $\sigma(\cdot)$ are aggregation and activation functions. Previous OSNC methods typically adopt a $C$-class closed-set classifier with threshold mechanism to detect OOD nodes. While tuning an optimal threshold to separate unknown from known is dataset-specific and time-consuming \cite{UAGA}. To get rid of threshold tuning, the proposed negMIX constructs a ($C$+1)-class open-set classifier $f(\cdot)$ by adding an additional class to represent the ``unknown'' class, so as to simultaneously accomplish OOD detection and ID classification in one step, as:%
\begin{equation}\label{eq.hat_y}
\displaystyle
\hat{y}_i=\operatorname{argmax}_{k\in\{1,...,C+1\}}p_{i,k}
\end{equation}%
where $p_{i,k}$ is the predicted probability of $v_i$ belonging to class $k$, and $\hat{y}_i$ is the predicted label. Given ground-truth labels $\boldsymbol{Y}_{\mathbb{L}}$ of training set, a cross-entropy loss is defined as: $\mathcal{L}_{ce}=-\frac1{|\mathcal{V}_{\mathbb{L}}|}\sum\nolimits_{i=1}\nolimits^{|\mathcal{V}_{\mathbb{L}}|}\log p_{i,y_i}$.

\subsection{Negative Mixup for OOD Generalization}
\label{sec:negMixup}
To enhance generalization, we propose negative Mixup to construct new pseudo-OOD/ID training samples, by negatively/positively mixing an original labeled ID node up with each selected potential OOD/ID node. 

\textbf{OOD Score with Neighborhood Aggregation.} To estimate the degree of a node belonging to ``unknown'' class, we compute the OOD score upon two signals, i.e., 1) the entropy over known classes and 2) the predicted probability of unknown class. A higher entropy indicates greater uncertainty in predicting a node as ID, suggesting that the node is more likely to be OOD. A higher unknown class probability directly reflects the higher confidence of a node belonging to ``unknown''. Additionally, recent studies have demonstrated that propagating the OOD score along graph structure can effectively enhance graph OOD detection performance \cite{gnnsafe,grasp,EMP}. While unlike previous methods just propagating either confidence \cite{grasp} or entropy \cite{EMP} alone, we compute the OOD score by jointly propagating entropy and unknown class probability among the neighborhood of each node $v_i$, as:%
\begin{equation}\label{eq.ood_score}
\displaystyle
OODscore_i=ent_i+p_{i,C+1}+\frac{1}{|\mathcal{N}_i|}\sum\nolimits_{v_j\in \mathcal{N}_i}(ent_j+p_{j,C+1})
\end{equation}%
where $ent_i$ is the normalized entropy of $v_i$ over $C$ known classes. Intuitively, if the neighbors of $v_i$ also exhibit high entropy over $C$ known classes and high confidence belonging to the ($C$+1)-th unknown class, then $v_i$ should be more likely to be OOD, consequently yielding a higher OOD score. 

\textbf{Highly Potential OOD/ID Selection via Clustering-then-Ranking.} Next, we select a set of highly potential OOD nodes $\mathcal{V}_{\mathbb{PO}}$ and a set of highly potential ID nodes $\mathcal{V}_{\mathbb{PI}}$ from the unlabeled test set $\mathcal{V}_{\mathbb{U}}$. A straightforward approach is to select top $\rho\%$ of nodes with the highest OOD score as $\mathcal{V}_{\mathbb{PO}}$ and those with the lowest OOD score as $\mathcal{V}_{\mathbb{PI}}$. However, this approach might contain noise in pseudo-labels. Intuitively, nodes near cluster centroids are more likely to be correctly predicted, making them more reliable candidates. Thus, instead of directly ranking OOD scores, we propose a \textbf{clustering-then-ranking} strategy, to firstly apply K-Means on the OOD scores of all unlabeled nodes to separate them into two clusters. Then, we select top $\rho\%$ of nodes closest to the high-score cluster centroid as $\mathcal{V}_{\mathbb{PO}}$ and those closest to the low-score cluster centroid as $\mathcal{V}_{\mathbb{PI}}$. This strategy can select more reliable potential OOD by favoring nodes that are both with high OOD score and representative of the OOD distribution. Then, a one-hot pseudo-label vector $\bar{\boldsymbol y}_i\in\mathbb{R}^{C+1}$ is assigned to each selected node. For $v_i^{\mathbb{PI}}\in\mathcal{V}_{\mathbb{PI}}$, the pseudo-label is the class with the highest probability among the pre-$C$ classes. For $v_i^{\mathbb{PO}}\in\mathcal{V}_{\mathbb{PO}}$, the pseudo-label is set to ``$C$+1''. Next, OOD score regularization is applied to encourage lower OOD score for labeled ID nodes, while higher OOD score for potential OOD nodes, as: 
\begin{equation}\label{eq.loss_os}
    \mathcal{L}_{oreg}=\frac{1}{|\mathcal{V}_{\mathbb{L}}|}\sum\nolimits_{i=1}\nolimits^{|\mathcal{V}_{\mathbb{L}}|}OOD score_{i}-\frac{1}{|\mathcal{V}_{\mathbb{PO}}|}\sum\nolimits_{j=1}\nolimits^{|\mathcal{V}_{\mathbb{PO}}|}OOD score_{j}
\end{equation}%


\textbf{Positive Mixup for Enhancing ID Generalization.} Mixup has been proven effective in improving the model’s generalization and reducing over-fitting, by creating new training samples with corresponding pseudo-labels through interpolating between different samples \cite{labelscarce3}. Following conventional Mixup, for each selected highly potential ID node $v_i^{\mathbb{PI}}\in\mathcal{V}_{\mathbb{PI}}$, we randomly select an original labeled ID node $v_j^{\mathbb{L}}\in\mathcal{V}_{\mathbb{L}}$, and combine them to construct a new mixed-up pseudo-ID training sample $(\boldsymbol{h}_i^{\mathbb{MI}},\boldsymbol{y}_i^{\mathbb{MI}})$ as:%
\begin{equation}\label{eq.mixup_pi}
\boldsymbol{h}_i^{\mathbb{MI}}
=\lambda\boldsymbol{h}_i^{\mathbb{PI}}+(1-\lambda)\boldsymbol{h}_j^{\mathbb{L}},
\quad
\boldsymbol{y}_i^{\mathbb{MI}}
=\lambda\bar{\boldsymbol{y}}_i^{\mathbb{PI}}+(1-\lambda)\boldsymbol{y}_j^{\mathbb{L}}
\end{equation}
where $\lambda\in[0,1]$ is a mixing coefficient randomly drawn from the Beta distribution. We feed the mixed-up ID embedding $\boldsymbol{h}_i^{\mathbb{MI}}$ into open-set classifier to obtain the predicted probability $\boldsymbol{p}_i^{\mathbb{MI}}$. The 
Mixup loss for training such mixed-up pseudo-ID samples is defined by a cross-entropy loss as:
\begin{align}\label{eq.l_pi}
\displaystyle 
\mathcal{L}_{pi}=-\frac1{|\mathcal{V}_{\mathbb{PI}}|}\sum\nolimits_{i=1}\nolimits^{|\mathcal{V}_{\mathbb{PI}}|}\sum\nolimits_{k=1}\nolimits^{C+1}y_{i,k}^{\mathbb{MI}}\log p_{i,k}^{\mathbb{MI}}
\end{align}%
The conventional Mixup between labeled ID and potential ID nodes can increase the diversity of training ID samples, thus alleviating over-fitting to limited labeled ID nodes and improving the model's generalization to unseen ID nodes.

\textbf{Whether Conventional Positive Mixup Feasible for OOD Generalization?} In OSNC, the ground-truth labels of OOD nodes are unavailable during training, making the open-set classifier lack direct supervision on the ($C$+1)-th class. In light of this, a straightforward approach is to adopt conventional Mixup to construct a new pseudo-OOD training sample by ``positively'' mixing a potential OOD node up with an original labeled ID node, similar to Eq. \ref{eq.mixup_pi}, as:
\begin{equation}\label{eq.conventional_mixup_po}
\boldsymbol{h}_i^{\mathbb{MO}}
=\lambda\boldsymbol{h}_i^{\mathbb{PO}}+(1-\lambda)\boldsymbol{h}_j^{\mathbb{L}},
\quad
\boldsymbol{y}_i^{\mathbb{MO}}
=\lambda\bar{\boldsymbol{y}}_i^{\mathbb{PO}}+(1-\lambda)\boldsymbol{y}_j^{\mathbb{L}}
\end{equation}
where the mixed-up pseudo-OOD embedding $\boldsymbol{h}_i^{\mathbb{MO}}$ would lie between the original labeled ID node $\boldsymbol{h}_j^{\mathbb{L}}$ and the potential OOD node $\boldsymbol{h}_i^{\mathbb{PO}}$, as illustrated in Figure \ref{fig.negmixup}a. Additionally, such mixed-up pseudo-OOD sample would be assigned with a soft positive label of both OOD class ($y_{i,C+1}^{\mathbb{MO}}=\lambda>0$) and ID class ($y_{i,y_j}^{\mathbb{MO}}=(1-\lambda)>0, y_j \in \mathcal{C}_\mathrm{\mathbb{L}}$).

To train the model on such positively mixed-up pseudo-OOD samples, a cross-entropy loss is applied to:

\begin{align}\label{eq.ce_po_1}
\mathcal{L}_{po}^{ce} 
&= -\frac{1}{|\mathcal{V}_{\mathbb{PO}}|} \sum\nolimits_{i=1}^{|\mathcal{V}_{\mathbb{PO}}|} \left(y_{i,C+1}^{\mathbb{MO}} \log p_{i,C+1}^{\mathbb{MO}} + y_{i,y_j}^{\mathbb{MO}} \log p_{i,y_j}^{\mathbb{MO}}\right)\notag
\\&=
-\frac{1}{|\mathcal{V}_{\mathbb{PO}}|}\sum\nolimits_{i=1}^{|\mathcal{V}_{\mathbb{PO}}|}\left(\lambda\log p_{i,C+1}^{\mathbb{MO}}+(1-\lambda)\log p_{i,y_j}^{\mathbb{MO}}\right)
\end{align}

We argue that \textbf{conventional positive Mixup would be problematic in enhancing OOD generalization in the open-set scenario}. This is because utilizing such positively mixed-up pseudo-OOD sample that lies between ID and OOD for model training would smooth the score function between ID and OOD, and unavoidably blur the ID/OOD boundary, thereby decaying both ID classification and OOD detection performance. We provide the theoretical justification in Theorem 1.

\label{Theorem 1}
\textbf{Theorem 1.}
    \textit{Given conventional positive Mixup between a selected potential OOD node $v_i^{\mathbb{PO}}\in\mathcal{V}_{\mathbb{PO}}$ and an original labeled ID node $v_j^{\mathbb{L}}\in\mathcal{V}_{\mathbb{L}}$ in Eq. \ref{eq.conventional_mixup_po}, minimizing the loss $-\lambda\log{p}_{i,C+1}^{\mathbb{MO}}$ would increase unknown class probability for both potential OOD node ($p_{i,C+1}^{\mathbb{PO}}\uparrow$) and original labeled ID node ($p_{j,C+1}^{\mathbb{L}}\uparrow$). Minimizing the loss $-(1-\lambda)\log p_{i,y_j}^{\mathbb{MO}}$ would increase known class probability for both original labeled ID node ($p_{j,y_j}^{\mathbb{L}}\uparrow$) and potential OOD node ($p_{i,y_j}^{\mathbb{PO}}\uparrow$).}

\textit{Remark}. The proof of Theorem 1 is provided in Appendix E. According to Theorem 1, optimizing the cross-entropy loss upon conventional positive Mixup would 1) undesirably increase unknown class probability for original labeled ID node, i.e., \textbf{increase the structural risk} of misclassifying ID nodes as OOD class, and 2) undesirably increase known class probability for potential OOD node, i.e., \textbf{increase the open space risk} of misclassifying OOD nodes as known class.

\textbf{Negative Mixup for Enhancing OOD Generalization.} As conventional positive Mixup is infeasible to enhance OOD generalization, we propose a novel negative Mixup specifically crafted for the open-set scenario, to \textbf{negatively combine selected potential OOD node with labeled ID node in an opposite direction, at both embeddings and labels}. Specifically, for each highly potential OOD node $v_i^{\mathbb{PO}}\in\mathcal{V}_{\mathbb{PO}}$, we randomly select a labeled ID node $v_j^{\mathbb{L}}\in\mathcal{V}_{\mathbb{L}}$ to reverse its embedding vector ($-\boldsymbol{h}_j^{\mathbb{L}}$) and assign it with a negative pseudo-label of the corresponding ID class ($-\boldsymbol{y}_j^{\mathbb{L}}$), finally constructing a new mixed-up pseudo-OOD training sample $(\boldsymbol{h}_i^{\mathbb{MO}},\boldsymbol{y}_i^{\mathbb{MO}})$ as: %
\begin{equation}\label{eq.mixup_po}
\boldsymbol{h}_i^{\mathbb{MO}}
=\lambda\boldsymbol{h}_i^{\mathbb{PO}}+(1-\lambda)(-\boldsymbol{h}_j^{\mathbb{L}}),
\quad
\boldsymbol{y}_i^{\mathbb{MO}}
=\lambda\bar{\boldsymbol{y}}_i^{\mathbb{PO}}+(1-\lambda)(-\boldsymbol{y}_j^{\mathbb{L}})
\end{equation}

Our negative Mixup is distinct from conventional positive Mixup at both label and embedding levels. At label-level, 1) a \textbf{positive label of unknown class} $y_{i,C+1}^{\mathbb{MO}}=\lambda>0$ indicates that ``the mixed-up pseudo-OOD training sample is likely to be OOD'', and 2) a \textbf{negative label of corresponding ID class} 
$y_{i,y_j}^{\mathbb{MO}}=-(1-\lambda)<0, y_j \in \mathcal{C}_\mathrm{\mathbb{L}}$ implies that ``the mixed-up pseudo-OOD training sample is unlikely to be ID''. At embedding-level, the negatively mixed-up pseudo-OOD sample would be positioned close to the OOD region while far away from the ID region, as illustrated in Figure \ref{fig.negmixup}b.

\textbf{Positive and Negative Learning Loss}. Unlike conventional Mixup always coupled with cross-entropy loss, a  positive and negative learning loss is devised to train the model upon negatively mixed-up pseudo-OOD samples. 
Firstly, positive learning loss is devised to optimize the output probability of mixed-up pseudo-OOD sample belonging to the positive label of unknown class to be 1 $({p}_{i,C+1}^{\mathbb{MO}}\to1)$, as:
\begin{align}\label{eq.loss_po_p}
    \displaystyle
\mathcal{L}_{po}^{pos}&= -\frac{1}{|\mathcal{V}_{\mathbb{PO}}|} \sum\nolimits_{i=1}^{|\mathcal{V}_{\mathbb{PO}}|} y_{i,C+1}^{\mathbb{MO}} \log p_{i,C+1}^{\mathbb{MO}}\notag \\&=-\frac{1}{|\mathcal{V}_{\mathbb{PO}}|}
\sum\nolimits_{i=1}^{|\mathcal{V}_{\mathbb{PO}}|}\lambda\log{p}_{i,C+1}^{\mathbb{MO}}
\end{align}%

The negative learning loss is carefully devised to optimize the output probability corresponding to the negative label of ID class to be 0 $({p}_{i,y_j}^{\mathbb{MO}}\to0, y_j \in \mathcal{C}_\mathrm{\mathbb{L}})$, as:
\begin{align}\label{eq.loss_po_n}
    \displaystyle
\mathcal{L}_{po}^{neg}&= -\frac{1}{|\mathcal{V}_{\mathbb{PO}}|} \sum\nolimits_{i=1}^{|\mathcal{V}_\mathbb{PO}|} \left| y_{i,y_j}^{\mathbb{MO}} \right| \log ( 1 - p_{i,y_j}^{\mathbb{MO}} )\notag \\
&=-\frac{1}{|\mathcal{V}_{\mathbb{PO}}|}
\sum\nolimits_{i=1}^{|\mathcal{V}_{\mathbb{PO}}|}(1-\lambda)\log(1-{p}_{i,y_j}^{\mathbb{MO}})
\end{align}%
Minimizing negative learning loss would effectively reduce the open space risk on misclassifying OOD nodes as ID class. The total positive and negative learning loss is: $\mathcal{L}_{po}=\mathcal{L}_{po}^{pos}+\mathcal{L}_{po}^{neg}$ .

Recall that each mixed-up pseudo-OOD sample is constructed by negatively combining an original labeled ID node and a selected potential OOD node in Eq. \ref{eq.mixup_po}. Therefore, optimizing the positive and negative learning loss $\mathcal{L}_{po}$ can also implicitly impact the output probability of such potential OOD node ($\boldsymbol{p}_{i}^{\mathbb{PO}}$) and original labeled ID node ($\boldsymbol{p}_{j}^{\mathbb{L}}$). We theoretically analyze such impact in Theorem 2.

\textbf{Theorem 2.}
    \textit{Given negative Mixup between a selected potential OOD node $v_i^{\mathbb{PO}}\in\mathcal{V}_{\mathbb{PO}}$ and an original labeled ID node $v_j^{\mathbb{L}}\in\mathcal{V}_{\mathbb{L}}$ in Eq. \ref{eq.mixup_po}, minimizing positive learning loss $-\lambda\log{p}_{i,C+1}^{\mathbb{MO}}$ would increase unknown class probability for potential OOD node ($p_{i,C+1}^{\mathbb{PO}}\uparrow$) and decrease unknown class probability for labeled ID node ($p_{j,C+1}^{\mathbb{L}}\downarrow$). Minimizing negative learning loss $-(1-\lambda)\log(1-{p}_{i,y_j}^{\mathbb{MO}})$ would decrease known class probability for potential OOD node ($p_{i,y_j}^{\mathbb{PO}}\downarrow$) and increase known class probability for labeled ID node ($p_{j,y_j}^{\mathbb{L}}\uparrow$).}

\textit{Remark}. The proof of Theorem 2 is provided in Appendix F. Supported by Theorem 2, the proposed negative Mixup can: 1) increase unknown class probability and decrease known class probability for potential OOD nodes, and 2) increase known class probability and decrease unknown class probability for labeled ID nodes, thus yielding clearer ID/OOD boundary to facilitate both ID classification and OOD detection. 

\captionsetup[subfigure]{font=scriptsize}
\begin{figure}[b]
\setlength{\belowcaptionskip}{9pt}
    \centering
    \begin{subfigure}[b]{0.49\linewidth}
        \centering
        \includegraphics[width=\linewidth]{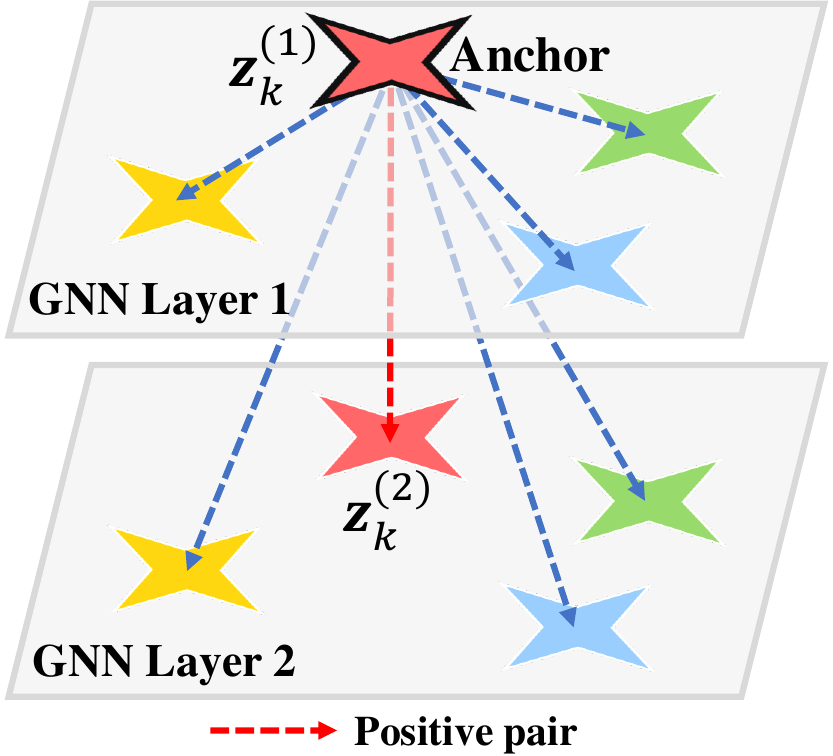} 
        \caption{Prototype-to-Prototype}
        \Description{The figure depicts prototype-to-prototype graph contrastive learning across GNN layers. Prototypes of the same class at different GNN layers are treated as augmented views and connected as positive pairs, while prototypes from different classes form negative pairs.}
        \label{cpgcl_sub1}
    \end{subfigure}
    \hfill
    \begin{subfigure}[b]{0.49\linewidth}
        \centering
        \includegraphics[width=\linewidth]{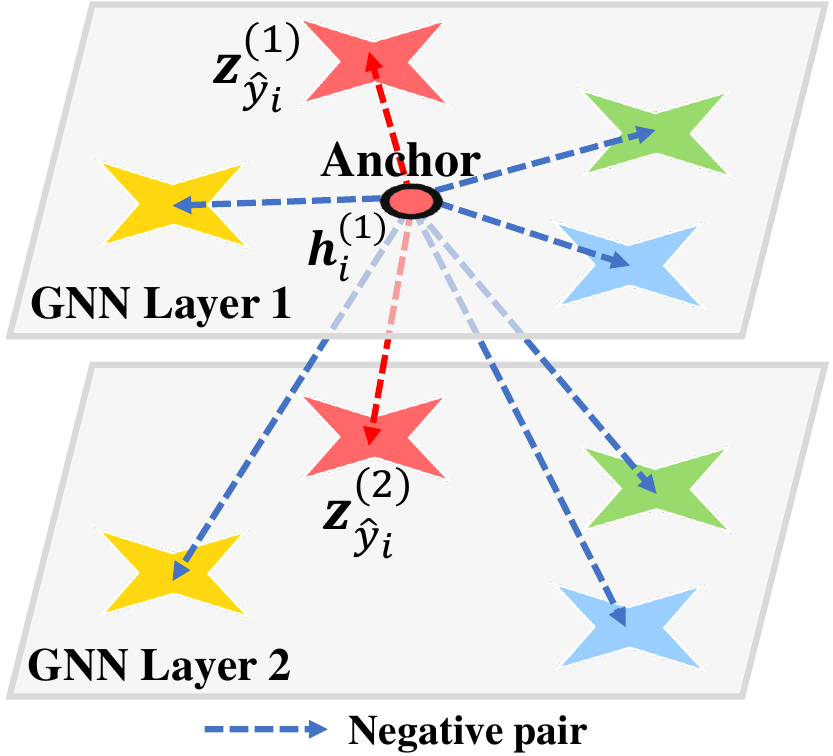} 
        \caption{Node-to-Prototype}
        \Description{The figure illustrates node-to-prototype graph contrastive learning across GNN layers. A node embedding at one GNN layer serves as the anchor and is pulled toward its corresponding class prototypes from both the same and different GNN layers, while being pushed away from prototypes of other classes.}
        \label{cpgcl_sub2}
    \end{subfigure}
    \caption{Illustration of Cross-layer GCL. Different colors correspond to different classes. Circles represent nodes and stars represent prototypes.}
    \label{cpgcl}
\end{figure}%

\subsection{Cross-layer Graph Contrastive Learning}
To promote intra-class compactness and inter-class separability, we devise a cross-layer GCL module with learnable graph augmentation and positive and negative pairs defined upon the class information. 

\textbf{GNN Layers as Learnable Graph Augmentation.}
It has been proven that good augmented views should be diverse while keeping task-relevant information intact \cite{goodview,NCLA,ROSEN}. In GCN-like models, earlier GNN layers capture local structure, while deeper layers capture global structure. That is, the \textbf{embeddings at different GNN layers indeed capture diverse neighborhood information with different topological distances} \cite{multilayersGnn}, which also exhibit strong correlation. Inspired by this, we consider different GNN layers as learnable augmented views, which not only avoids improper modification of task-relevant information but also eliminates handcraft graph augmentations.

Next, we devise a cross-layer GCL scheme to maximize the prototypical mutual information across different GNN layers, at prototype-to-prototype and node-to-prototype levels. The  prototypes are computed upon the class labels of nodes, where unlabeled nodes use the pseudo-labels $\hat{y}_i$ in Eq. \ref{eq.hat_y}, and labeled nodes use the ground-truth labels $\boldsymbol{Y}_\mathbb{L}$. Specifically, for each class $k$ at layer $l$, the class prototype is computed as the mean of embeddings of all nodes at layer $l$ with the class label of $k$: 
    $\boldsymbol{z}_k^{(l)} = \bigl(\sum_{v_i\in\mathcal{V}}\vmathbb{1}\{\hat{y}_i=k\}\cdot\boldsymbol{h}_i^{(l)}\bigr)/\sum_{v_i\in\mathcal{V}}\vmathbb{1}\{\hat{y}_i=k\}$, where $\vmathbb{1}\{\cdot\}$ is an indicator function. 

\textbf{Prototype-to-Prototype Graph Contrastive Learning} aims to maximize the mutual information between the same class prototypes across neighborhood with diverse topological distances (i.e. different GNN layers). As illustrated in Figure \ref{cpgcl_sub1}, let GNN layer 1 and 2 be two augmented views, given the prototype of class $k$ at GNN layer 1 as the anchor ($\boldsymbol{z}_k^{(1)}$), it forms positive pair with the same class prototype at different GNN layer ($\boldsymbol{z}_k^{(2)}$), and negative pairs with different class prototypes within the same GNN layer and from different GNN layer. Accordingly, the prototype-to-prototype contrastive loss associated with the anchor $\boldsymbol{z}_k^{(1)}$ between GNN layer 1 and 2 is defined by the modified NT-Xent loss \cite{SimCLR} as: 
\begin{align}\label{eq.loss_p2p1}
\displaystyle
&\resizebox{0.7\linewidth}{!}{$
\displaystyle
\quad \quad \quad \quad \quad  \quad  \quad \quad \quad \quad \quad
\mathcal{L}_{p2p}(\boldsymbol{z}_k^{(1)})=
\quad \quad \quad \quad \quad \notag$}
\\
&\resizebox{0.9\linewidth}{!}{$
\displaystyle
-\mathrm{log}\frac{e^{\mathrm{\theta}(\boldsymbol{z}_k^{(1)},\boldsymbol{z}_k^{(2)})/\tau}}{e^{\mathrm{\theta}(\boldsymbol{z}_k^{(1)},\boldsymbol{z}_k^{(2)})/\tau}+\sum_{i=1,i\neq k}^{C+1}(e^{\mathrm{\theta}(\boldsymbol{z}_k^{(1)},\boldsymbol{z}_i^{(1)})/\tau}+e^{\mathrm{\theta}(\boldsymbol{z}_k^{(1)},\boldsymbol{z}_i^{(2)})/\tau})}$}
\end{align}%
where $\mathrm{\theta}(\cdot)$ denotes the cosine similarity, and $\tau$ is the temperature parameter. Minimizing Eq. \ref{eq.loss_p2p1} pulls the same class prototypes from different GNN layers closer, while pushing different class prototypes farther away. Due to the symmetry of different GNN layers, given $\boldsymbol{z}_k^{(2)}$ as the anchor, the contrastive loss $\mathcal{L} _{p2p}( \boldsymbol{z}_k^{(2)})$ can be similarly computed. The contrastive loss between GNN layer l and 2 is then averaged by selecting each class prototype as the anchor: $\mathcal{L}_{p2p}(\boldsymbol{Z}^{(1)},\boldsymbol{Z}^{(2)})=\frac{1}{2(C+1)}\sum\nolimits_{k=1}^{C+1}[\mathcal{L}_{p2p}(\boldsymbol{z}_k^{(1)})+\mathcal{L}_{p2p}\big(\boldsymbol{z}_k^{(2)})]$. When more than two GNN layers are available, a GNN layer $p$ is randomly selected as the pivot, and the total prototype-to-prototype contrastive loss is formed by averaging between the pivot layer $p$ and each other layer $q$ as: $\mathcal{L}_{p2p}=\frac1L\sum\nolimits_{q=1,q\neq p}^L\mathcal{L}_{p2p}(\boldsymbol{Z}^{(p)},\boldsymbol{Z}^{(q)})$.%

\textbf{Node-to-Prototype Graph Contrastive Learning} aims to map each node close to its corresponding class prototypes across all GNN layers, while distinguishing it from other class prototypes. As illustrated in Figure \ref{cpgcl_sub2}, selecting node $v_i$ at GNN layer 1 as the anchor ($\boldsymbol{h}_i^{(1)}$), the positive samples consist of the prototypes corresponding to the pseudo-label $\hat{y}_i$ within the same GNN layer ($\boldsymbol{z}_{\hat{y}_i}^{(1)}$) and from different GNN layer ($\boldsymbol{z}_{\hat{y}_i}^{(2)}$). While negative samples are different class prototypes at different GNN layer. The node-to-prototype contrastive loss associated with the anchor node $\boldsymbol{h}_i^{(1)}$ is defined by the modified NT-Xent loss as:%
\begin{equation}\label{eq.loss_n2p}
\displaystyle
\mathcal{L}_{n2p}(\boldsymbol{h}_{i} ^{(1)})=-\mathrm{log}\frac{(e^{\mathrm{\theta}(\boldsymbol{h}_{i}^{(1)},\boldsymbol{z}_{\hat{y}_{i}}^{(1)})/\tau}+e^{\mathrm{\theta}(\boldsymbol{h}_{i}^{(1)},\boldsymbol{z}_{\hat{y}_{i}}^{(2)})/\tau})/2}{\sum_{k=1}^{C+1}(e^{\mathrm{\theta}(\boldsymbol{h}_{i}^{(1)},\boldsymbol{z}_{k}^{(1)})/\tau}+e^{\mathrm{\theta}(\boldsymbol{h}_{i}^{(1)},\boldsymbol{z}_{k}^{(2)})/\tau})}
\end{equation}%
Selecting each node as the anchor, the total node-to-prototype contrastive loss $\mathcal{L}_{n2p}$ across different GNN layers can be computed similarly to $\mathcal{L}_{p2p}$. The total GCL loss is $\mathcal{L}_{gcl}= \mathcal{L}_{p2p}+ \mathcal{L}_{n2p}$. 

\textbf{The total loss of negMIX} is defined as:%
\begin{equation}\label{eq.total_loss}
\displaystyle
\mathcal{L}=\mathcal{L}_{ce}+\gamma\mathcal{L}_{oreg}+\eta\mathcal{L}_{pi}+\delta\mathcal{L}_{po}+\beta\mathcal{L}_{gcl}
\end{equation}%
where $\gamma$, $\eta,\delta$ and $\beta$ are trade-off parameters. 

\textbf{Complexity Analysis.} The time complexity of GAT encoder is linear to the number of nodes and edges. The time complexity of negative Mixup is $O((|\mathcal{V}_\mathbb{PI}|+|\mathcal{V}_\mathbb{PO}|)CF')$, where $|\mathcal{V}_\mathbb{PI}|$ and $|\mathcal{V}_\mathbb{PO}|$ are the number of selected potential ID and OOD nodes, $F'$ is the embedding dimension, and $C$ is the number of known classes. The time complexity of cross-layer Prototype-to-Prototype GCL is $O(C^2F'L)$, and of cross-layer Node-to-Prototype GCL is $O(NCF'L)$, where $L$ is the number of GNN layers and $N$ is the number of nodes. Note that $F'$, $C$ and $L$ are much smaller than $N$, thus the time complexity of cross-layer GCL is linear to the number of nodes. The overall time complexity of negMIX is linear to the number of nodes and edges.

\section{Experiments}
\label{sec:Experiments}
\textbf{Datasets.} Cora, Citeseer, PubMed \cite{coraciteseerpubmed} are citation networks, AmazonComputers and AmazonPhoto \cite{shchur2019pitfalls} are co-purchasing networks, CoauthorCS \cite{shchur2019pitfalls} is a coauthor network, Wiki-CS \cite{wikics} is a hyperlink network, and Arxiv \cite{Arxiv} is a large-scale citation network. 

\textbf{Baselines and Metrics.} We competed against eight SOTA OSNC methods, including \textbf{OpenWGL} \cite{OpenWGL}, \textbf{OODGAT} \cite{OODGAT}, \textbf{GNNSAFE++} \cite{gnnsafe}, \textbf{G$^2$Pxy} \cite{G2Pxy}, \textbf{EMP} \cite{EMP}, \textbf{GRASP} \cite{grasp}, \textbf{NODESAFE++} \cite{nodesafe} and \textbf{GOLD} \cite{GOLD}. We adopt $(C+1)$-class Accuracy, Macro-F1, AUROC and FPR@95 as metrics, following \cite{OODGAT,EMP}. Higher Accuracy, Macro-F1 and AUROC, while lower FPR@95 represent better performance.

\textbf{Implementation Details.}
The proposed negMIX\footnote{The code of negMIX is released at https://github.com/JunweiGong/negMIX.} was implemented in PyTorch 2.2.1 and PyTorch Geometric 2.4.0. The negMIX was trained by Adam optimizer with a learning rate of 1e-2. The negMIX was run with 1000 epochs and the model with lowest validation loss was chosen for test. The ratio $\rho$\%  of potential OOD/ID nodes selected for Mixup was set to 10\%. The trade-off parameters $\gamma$ and $\eta$ were selected from \{0.1, 1\}, $\delta$ and $\beta$ were selected from \{0.1, 1, 10\}. 


\subsection{Open-set Node Classification Performance}
To simulate OSNC scenario, for each dataset, we treat the first half of classes as ID and the remaining classes as OOD. Then, we randomly sampled 10\% of ID nodes for training, 10\% of ID nodes for validation, and 80\% of ID nodes and all OOD nodes for test. The average scores and standard deviations over 10 random splits are reported in Table \ref{mainresults}. We have key observations as follows:

Overall, the proposed negMIX demonstrates significant outperformance over SOTA OSNC baselines. Compared to the competitive baseline NODESAFE++, negMIX still achieves significant performance gains, improving accuracy by 1.28\% to 64.82\% and Macro-F1 by 1.38\% to 70.65\% across eight datasets. Moreover, for OOD detection, compared to the second-best results across eight datasets, negMIX improves AUROC by 0.36\% to 4.87\% and substantially reduces FPR@95 by 2.00\% to 57.54\%. This highlights the superior OOD detection performance of our negMIX over SOTA methods. The outperformance of negMIX might benefit from two novel designs: 1) negative Mixup to improve generalization to unseen ID and OOD nodes and yield clearer boundary between ID and OOD, and 2) cross-layer GCL to enhance intra-class compactness and inter-class separability. %

\begin{table*}
\caption{OSNC results on eight benchmark datasets. $\uparrow$ ($\downarrow$) indicates larger (smaller) values are better. The best and second-best results are highlighted with \textbf{bold} and \underline{underline}. OOM means Out-Of-Memory.}   
\label{mainresults}
\centering
\renewcommand{\arraystretch}{1.0}
\setlength{\tabcolsep}{3.5mm}{
    \resizebox{0.95\linewidth}{!}{
    \begin{tabular*}{\linewidth}{ccccc}
        \toprule
        \multirow{2}{*}{{\textbf{Method}}}
                        &\multicolumn{4}{c}{\textbf{Accuracy $\uparrow$ / Macro-F1 $\uparrow$} }\\
        &{\textbf{Cora}}&{\textbf{Citeseer}}&{\textbf{PubMed}}&{\textbf{AmazonComputers}}

\\
      \hline
      {OpenWGL}&69.07±1.70/67.99±2.12&68.37±0.98/47.44±2.39&74.96±0.77/73.60±0.81&76.19±0.50/53.54±1.53
\\
      {OODGAT}&71.69±4.63/75.20±2.48&67.70±4.30/54.53±2.09&71.58±3.45/75.73±1.24&85.65±0.50/87.98±0.98
\\
      {G$^2$Pxy}&70.37±2.70/66.21±4.00&69.43±6.39/50.76±2.54&62.31±1.62/61.49±1.57&66.68±3.49/43.32±3.48
\\
{GNNSAFE++}&77.68±2.81/77.04±2.50&71.58±2.33/55.79±4.50&77.72±2.16/77.49±2.14&84.05±2.59/85.05±2.04
\\
{EMP}&75.45±1.69/71.89±2.57&70.66±0.96/47.75±2.85&73.66±4.14/73.30±4.85&84.62±0.33/62.86±2.23
\\
{GRASP}&78.89±2.20/77.17±2.34&71.36±1.88/55.25±2.27&78.44±3.99/77.30±4.49&85.69±1.99/82.56±3.43

\\
{NODESAFE++}&\underline{79.74±2.04}/\underline{78.81±2.22}&\underline{72.90±1.59}/\underline{57.70±1.86}&\underline{78.76±2.32}/\underline{78.18±2.09}&\underline{88.49±0.69}/\underline{88.66±0.90}

\\
{GOLD}&76.74±2.23/76.37±1.93&69.45±2.90/55.22±2.41&74.11±2.43/73.68±2.25&83.39±2.68/84.88±2.47

\\
{\textbf{negMIX(Ours)}}&\textbf{84.29±0.82}/\textbf{82.78±0.87}&\textbf{76.47±0.95}/\textbf{59.25±1.67}&\textbf{84.21±0.71}/\textbf{83.35±0.70}&\textbf{90.77±0.45}/\textbf{89.88±1.49}\\
\hline
      &\textbf{AmazonPhoto}&  \textbf{CoauthorCS}&\textbf{ WikiCS}&\textbf{Arxiv}\\
      \hline
      {OpenWGL}&84.04±2.91/78.65±3.80& 85.09±0.51/72.94±2.39&73.74±0.80/69.73±2.30&OOM
\\
      {OODGAT}&86.30±4.36/88.44±3.45& 85.19±2.77/82.95±2.31&68.78±3.38/\underline{80.85±0.78}&45.32±1.14/23.81±0.34
\\
      {G$^2$Pxy}&73.75±2.92/64.54±5.30& 81.22±2.48/62.68±7.89&61.78±9.06/55.12±13.62&32.60±9.99/7.17±2.52
\\
{GNNSAFE++}&89.11±1.68/89.77±1.29& 88.02±1.09/83.74±0.97& 71.32±2.77/73.28±1.92&41.74±2.97/23.52±0.89
\\
{EMP}&92.53±0.47/89.56±0.57& \underline{90.69±0.47}/\underline{84.55±1.06}& 77.18±1.70/76.47±1.39&OOM
\\
{GRASP}&90.74±1.76/88.47±1.39&87.86±0.69/82.82±0.48&\underline{78.98±1.03}/78.47±0.86&\underline{58.76±1.42}/\textbf{33.38±1.29}

\\
{NODESAFE++}&\underline{93.31±0.43}/\underline{90.84±0.97}&88.39±0.59/82.31±1.18&77.51±0.73/71.41±3.25&36.98±0.45/17.72±0.92

\\
{GOLD}&91.92±1.35/89.59±1.55&88.95±0.88/83.86±0.75&75.61±1.91/76.82±1.38&OOM

\\

 {\textbf{negMIX  (Ours)}}&\textbf{94.50±0.23}/\textbf{92.49±0.35}&\textbf{92.03±0.33}/\textbf{84.76±1.43}&\textbf{82.17±0.94}/\textbf{81.19±4.56}&\textbf{60.95±0.57}/\underline{30.24±0.97}
 \\
        \midrule
        \multirow{2}{*}{\textbf{Method}}
        &\multicolumn{4}{c}{\textbf{AUROC $\uparrow$ / FPR@95 $\downarrow$}} \\&\textbf{Cora}&\textbf{Citeseer}&\textbf{PubMed}&\textbf{AmazonComputers}
                
\\
      \hline
      {OpenWGL}&81.21±1.75/64.51±5.43&75.09±2.12/73.02±2.80&81.46±0.90/66.22±2.92&90.83±0.72/33.88±2.30
\\
      {OODGAT}&88.94±1.79/41.90±4.85&80.72±3.22/62.97±5.32&82.45±1.91/65.59±6.55&95.48±0.67/24.22±3.96
\\
      { G$^2$Pxy}&85.40±1.74/48.91±7.50&80.37±1.68/60.55±5.42&78.40±1.72/70.48±3.60&82.12±2.07/48.38±3.66
\\
                 { GNNSAFE++}&90.38±3.23/42.53±13.29&83.90±2.00/55.98±4.44&86.62±2.02/62.87±4.88&92.28±1.55/27.06±7.48
\\
                       { EMP}&89.97±1.29/43.07±6.00&79.68±1.31/64.68±2.50&81.54±4.88/62.70±5.16&\underline{96.02±0.22}/21.09±0.84
\\
{ GRASP}&91.41±1.33/\underline{36.18±6.91}&80.16±1.94/58.48±3.93&87.17±3.03/62.83±1.79&94.31±0.85/19.49±10.04
\\
{ NODESAFE++}&
91.80±2.27/37.86±7.61&\underline{84.99±2.24}/\underline{52.76±4.86}&\underline{87.59±2.33}/59.29±5.84&94.62±0.54/\underline{14.62±3.34}

\\
{ GOLD}&\underline{92.07±1.86}/50.01±6.99&73.65±4.73/64.94±9.28&82.04±2.75/\underline{52.92±8.66}&95.31±0.89/43.84±9.01

\\

 {\textbf{ negMIX  (Ours)}}&\textbf{94.50±0.87}/\textbf{29.81±6.52}&\textbf{86.50±1.18}/\textbf{49.88±3.91}&\textbf{91.86±0.35}/\textbf{45.97±2.35}&\textbf{97.38±0.10}/\textbf{12.60±1.22}\\
        \hline
&\textbf{AmazonPhoto}&  \textbf{CoauthorCS}&\textbf{ WikiCS}&\textbf{Arxiv}
\\
      \hline
      { OpenWGL}&92.81±3.39/29.65±9.02& 94.22±0.33/30.64±1.49&87.67±0.81/50.79±2.97&OOM
\\
      { OODGAT}&96.27±2.31/21.38±16.08& 94.18±2.59/35.34±16.01&\underline{92.11±0.54}/38.28±2.16&65.73±0.46/76.88±0.81
\\
      { G$^2$Pxy}&86.69±1.70/54.36±4.45& 90.42±0.62/43.14±4.05&74.04±2.38/65.11±5.60&59.09±2.34/89.18±2.21
\\
                 { GNNSAFE++}&95.55±1.03/14.33±4.74& 96.33±0.75/20.39±4.89& 83.00±2.44/54.32±4.59&64.86±1.45/80.73±1.08
\\
                       { EMP}&\underline{98.47±0.11}/7.11±1.19& \underline{97.68±0.21}/\underline{10.25±1.35}& 89.85±2.04/46.76±11.56&OOM
\\
{ GRASP}&96.19±0.44/12.33±6.46&96.04±0.20/19.25±1.33&89.82±0.73/\underline{35.83±2.01}&\underline{78.69±0.62}/\underline{65.5±2.62}
\\
{ NODESAFE++}&97.01±0.16/\underline{4.64±0.35}&96.57±0.39/16.50±1.77&89.15±0.61/39.08±3.07&59.35±0.51/79.99±0.73

\\
{ GOLD}&96.43±0.76/7.77±3.08&95.06±0.61/14.76±1.55&90.73±1.41/66.67±4.89&OOM

\\

 { \textbf{negMIX (Ours)}}&\textbf{99.05±0.09}/\textbf{1.97±0.23}&\textbf{98.03±0.14}/\textbf{8.31±0.56}&\textbf{93.31±1.20}/\textbf{35.06±8.23}&\textbf{82.34±0.27}/\textbf{64.19±0.44}\\
        \hline
        \end{tabular*}}
        }
\end{table*}

\begin{figure*}[htbp]
    \centering
    \includegraphics[width=\linewidth]{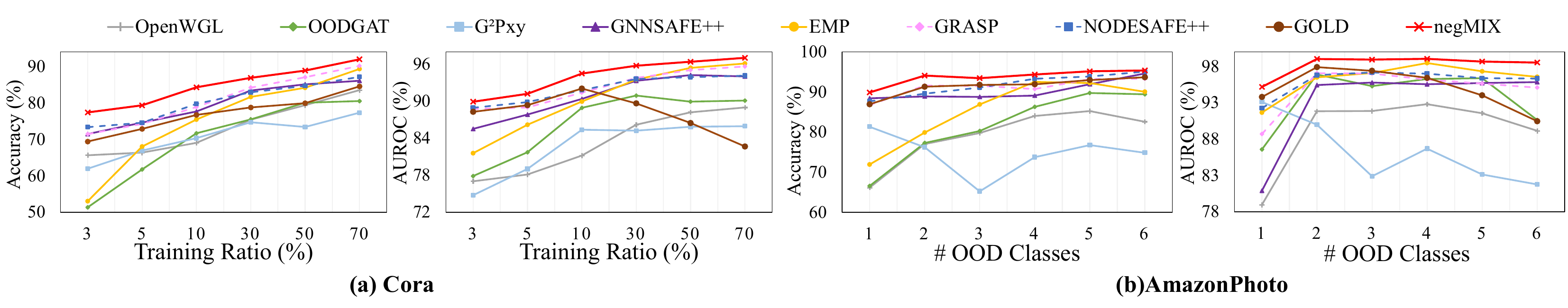} 
    \caption{OSNC performance under (a) different training ratios and (b) different numbers of OOD classes.}
    \Description{The figure compares negMIX with eight baselines across different training ratios and numbers of OOD classes, illustrating its stable performance under different levels of label scarcity and OOD diversity.}
    \label{fig:cross_column_figure}
\end{figure*}

\textbf{Performance under Different Training Ratios.} As shown in Figure \ref{fig:cross_column_figure}a, the proposed negMIX consistently outperforms SOTA baselines under different training ratios from 3\% to 70\%, highlighting the robustness of negMIX across different levels of label scarcity.

\textbf{Performance under Different Numbers of OOD Classes.} As shown in Figure \ref{fig:cross_column_figure}b, the performance of several OSNC baselines are sensitive to the number of OOD classes. However, our negMIX remains more robust and consistently outperforms SOTA baselines under different numbers of OOD classes.

\begin{figure}
\setlength{\abovecaptionskip}{3pt}
\setlength{\belowcaptionskip}{-1pt}
\begin{minipage}{\linewidth}
    \centering
    \begin{subfigure}{0.32\linewidth}
        \centering
        \includegraphics[width=\linewidth]{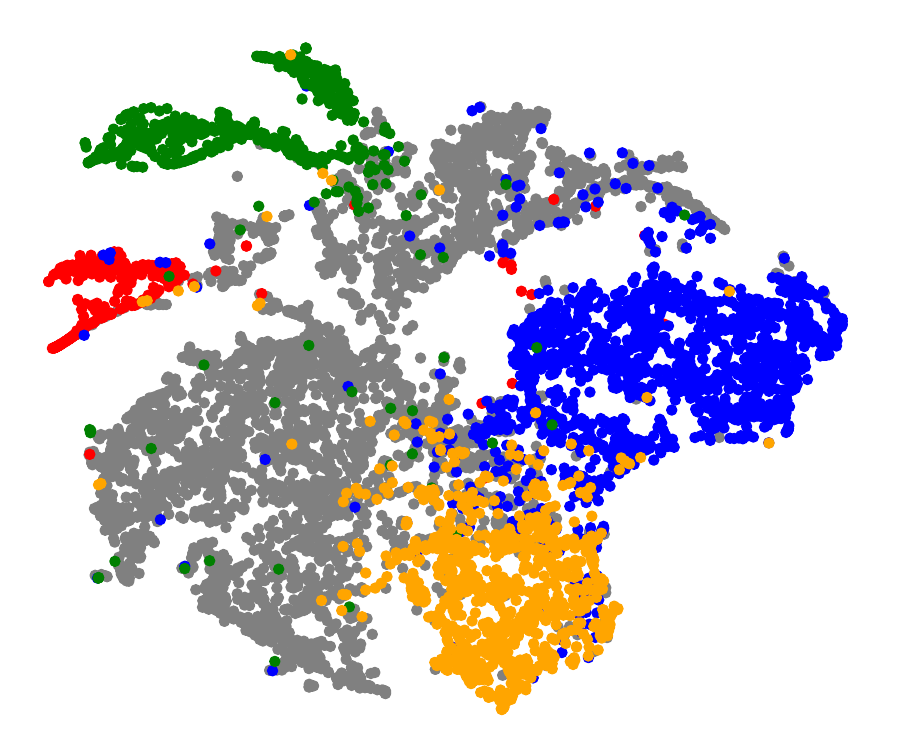} 
        \caption{GRASP}
        \Description{The Visualization of embeddings of GRASP on AmazonPhoto.}
        \label{tsne_grasp}
    \end{subfigure}
    \begin{subfigure}{0.32\linewidth}
        \centering
        \includegraphics[width=\linewidth]{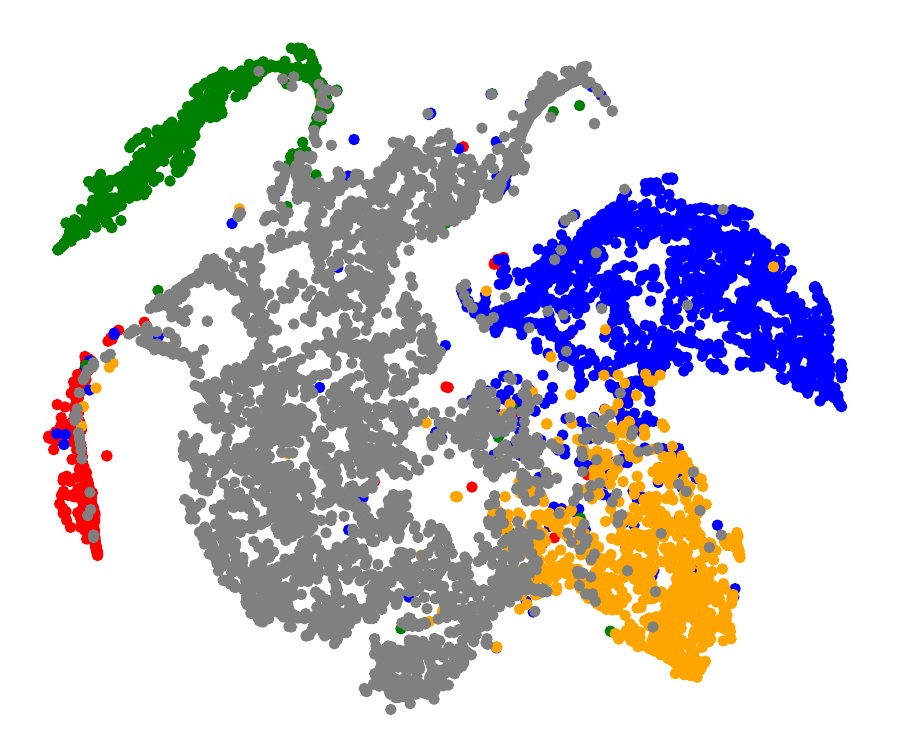} 
        \caption{NODESAFE++}
        \Description{The Visualization of embeddings of NODESAFE++ on AmazonPhoto.}
        \label{tsne_nodesafe}   
    \end{subfigure}
    \begin{subfigure}{0.32\linewidth}
        \centering
        \includegraphics[width=\linewidth]{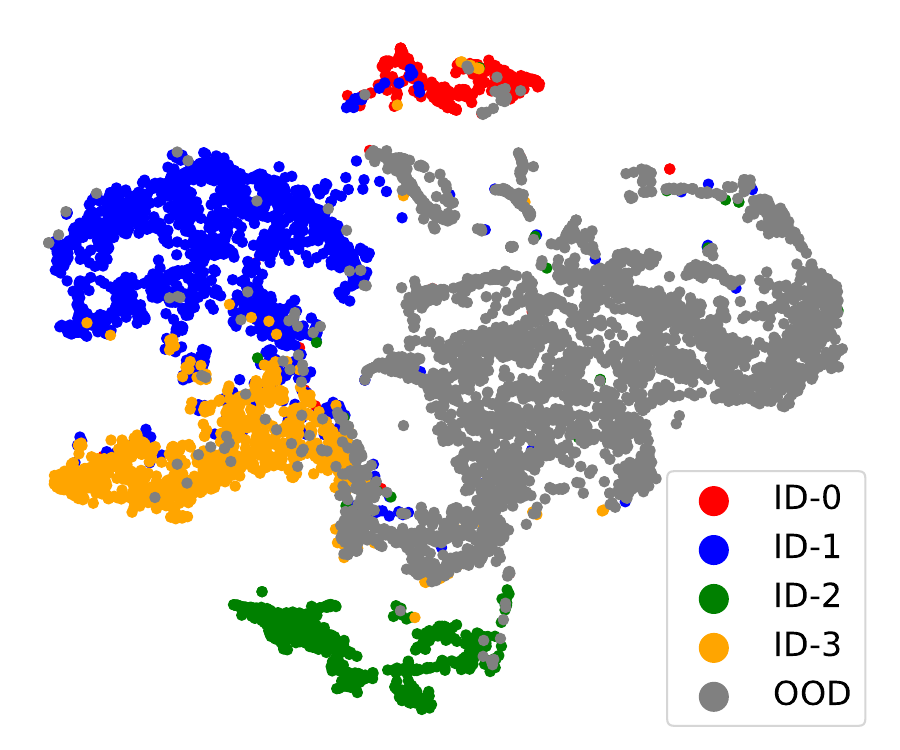} 
        \caption{negMIX (Ours)}
        \Description{The Visualization of embeddings of negMIX on AmazonPhoto.}
    \label{tsne_negMIX}
    \end{subfigure}
    \caption{Visualization of embeddings on AmazonPhoto. Grey color denotes the OOD class.}
    \label{fig.tsne}
\end{minipage}

\begin{minipage}[t]{\linewidth}
\setlength{\abovecaptionskip}{3pt}
\setlength{\belowcaptionskip}{-1pt}
    \centering
    \begin{subfigure}[b]{0.45\linewidth}
        \centering
        \includegraphics[width=\linewidth]{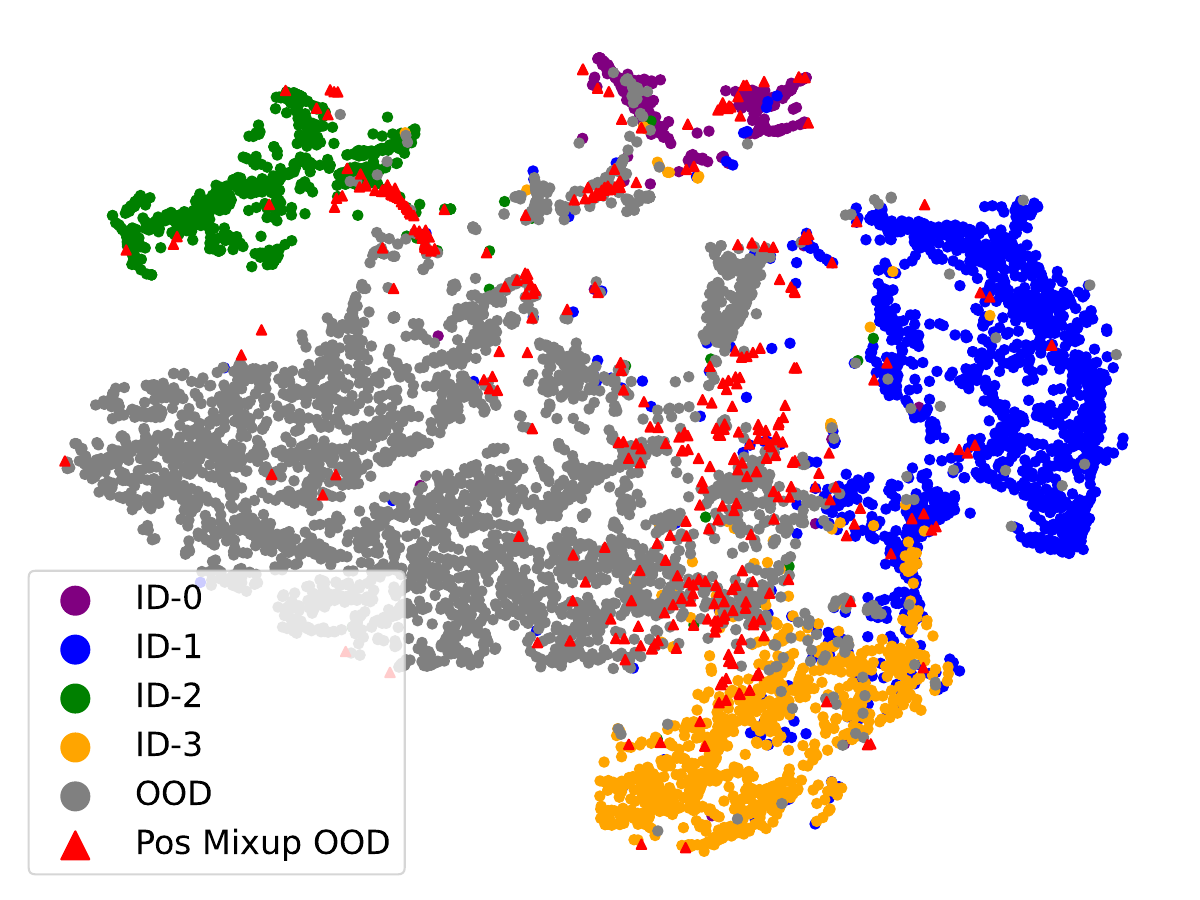} 
        \caption{Conventional Positive Mixup}
        \Description{The Visualization of embeddings of Conventional Positive Mixup.}
        \label{subfig_conmix}
    \end{subfigure}
    \begin{subfigure}[b]{0.45\linewidth}
        \centering
        \includegraphics[width=\linewidth]{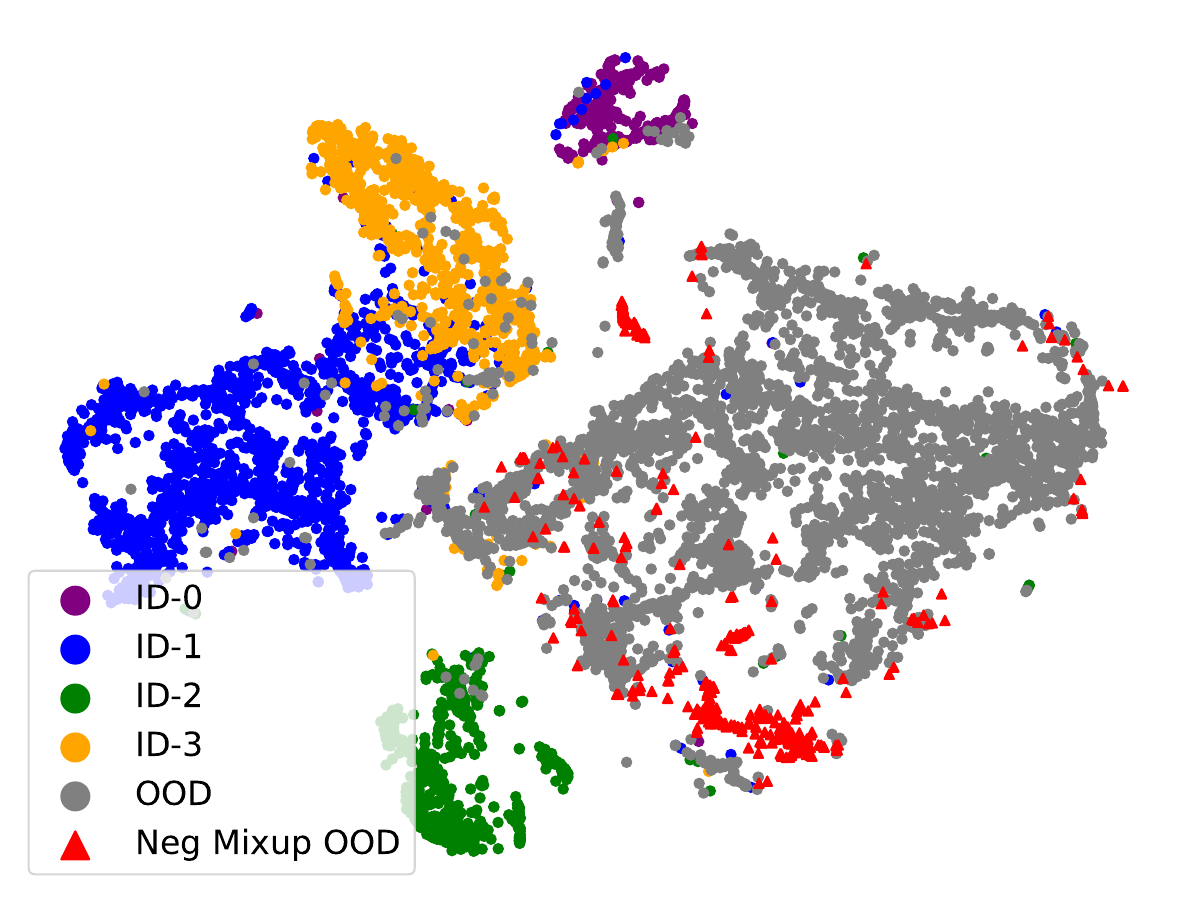} 
        \caption{Negative Mixup}
        \Description{The Visualization of embeddings of negative Mixup.}
        \label{subfig_negmix}   
    \end{subfigure}
    \caption{Visualization comparison between conventional positive Mixup and our negative Mixup on AmazonPhoto. Red triangles denote mixed-up pseudo-OOD samples.
}
    \label{fig.Visual_negmix_conmix}
\end{minipage}

\end{figure}

\begin{figure}
\setlength{\abovecaptionskip}{3pt}
\setlength{\belowcaptionskip}{-1pt}
    \centering
    \begin{subfigure}{0.45\linewidth}
        \centering
        \includegraphics[width=\linewidth]{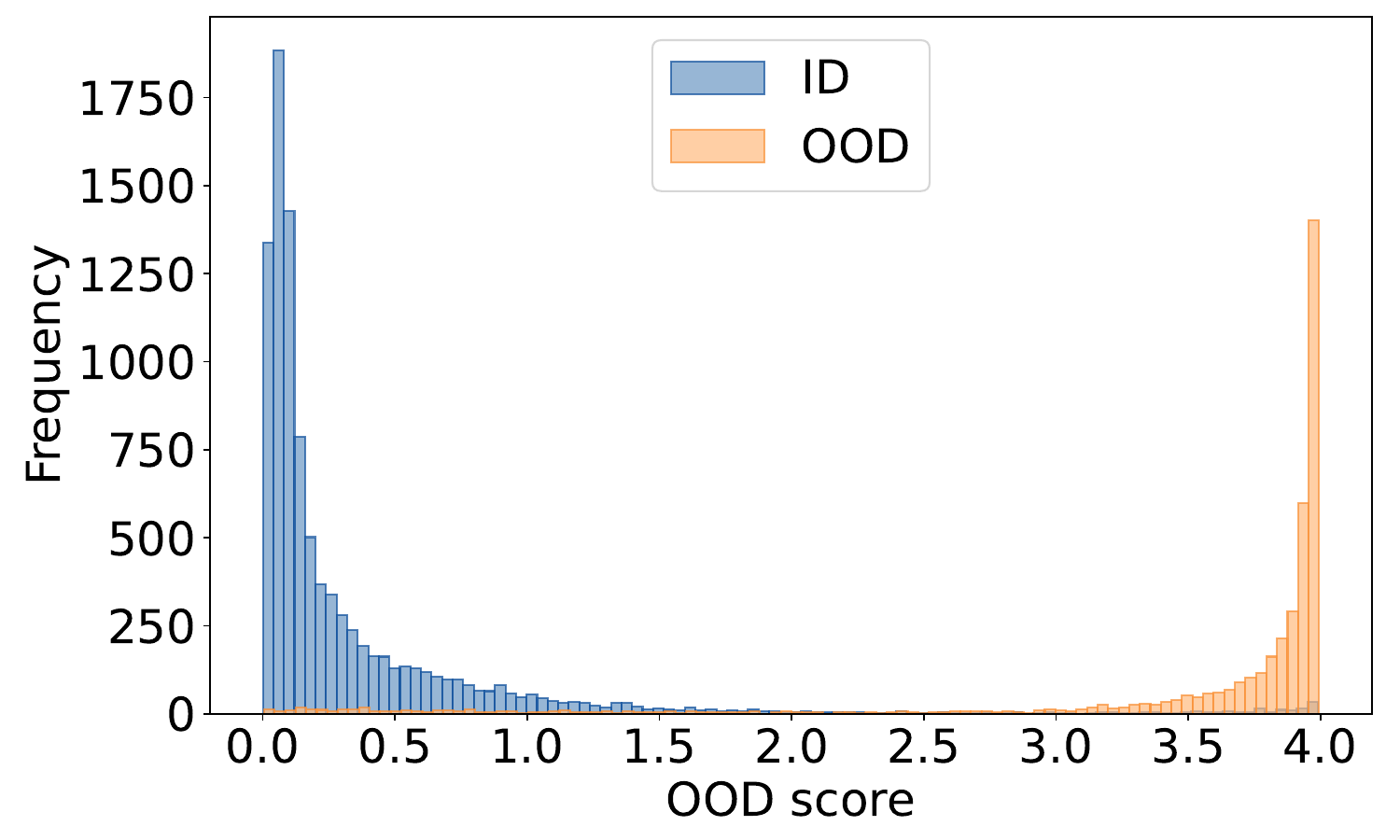} 
        \caption{AmazonComputers}
        \Description{The visualization of OOD Score on AmazonComputers.}
        \label{fig:subfig.va}
    \end{subfigure}
    \begin{subfigure}{0.45\linewidth}
        \centering
        \includegraphics[width=\linewidth]{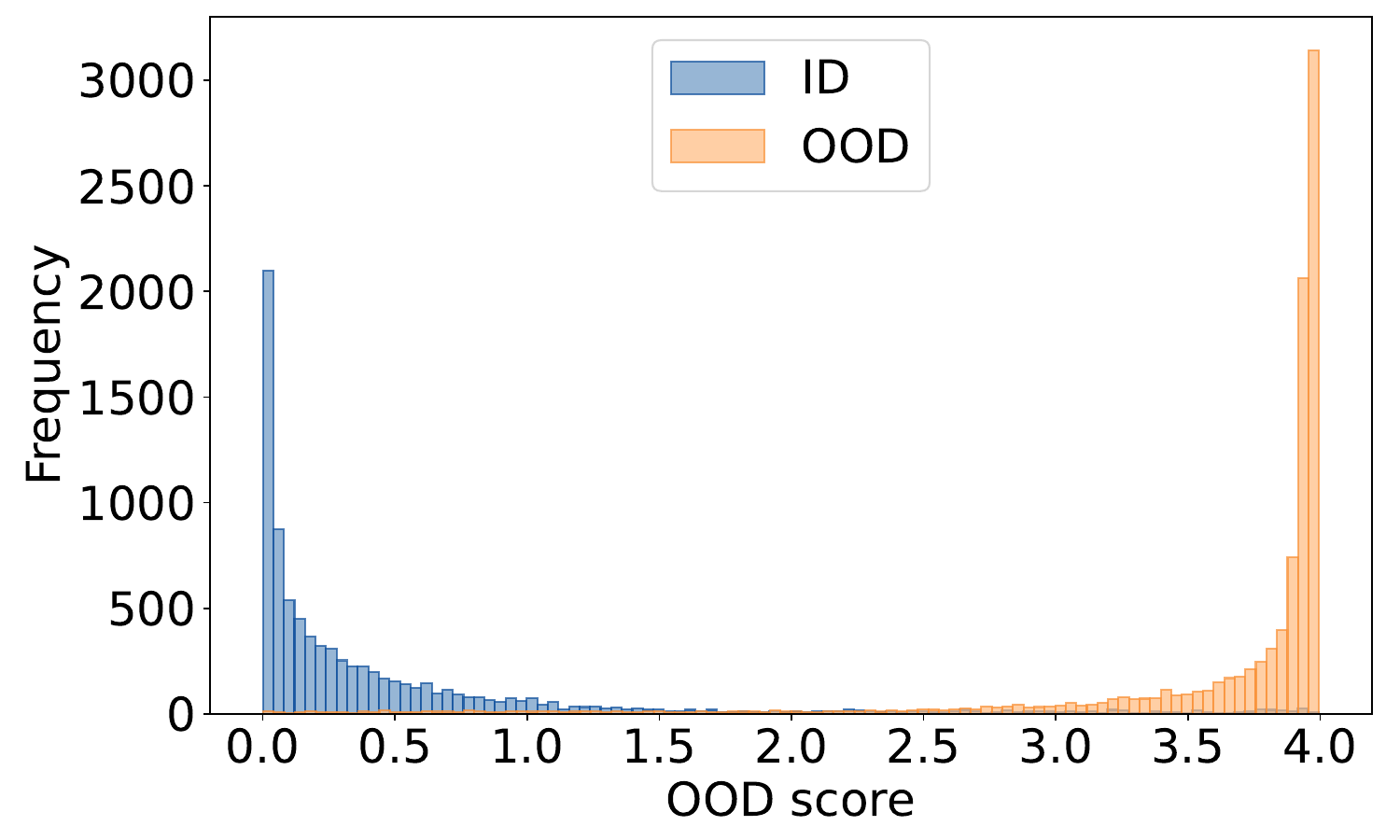} 
        \caption{CoauthorCS}
        \Description{The visualization of OOD Score on CoauthorCS.}
        \label{fig:subfig.vb}   
    \end{subfigure}
    \caption{Visualization of OOD score distribution.}
    \label{fig.Visual}
\end{figure}

\subsection{Analysis of negMIX}
\textbf{Ablation Study}. As shown in Table \ref{tab:ablation_study}, 1) without positive Mixup to construct new pseudo-ID training samples would lead to worse performance, highlighting the effectiveness of conventional Mixup on improving generalization to unseen ID nodes. 2) Replacing negative Mixup with conventional positive Mixup to construct new pseudo-OOD training samples would consistently degrade the performance. This is because conventional Mixup tends to blur the ID and OOD boundary in open-set scenario. 3) Directly using selected potential OOD nodes for training, instead of negative Mixup, the OOD detection performance would drop considerably, e.g., increasing FPR@95 by 495\% on AmazonPhoto. This highlights that our negative Mixup can construct more diverse OOD training samples to enhance the model's generalization to unseen OOD nodes. 4) Eliminating either positive or negative learning loss would significantly weaken the performance. This perfectly aligns with our theoretical analysis in Theorem 2, i.e., jointly optimizing such two losses would facilitate clearer ID/OOD boundary. 5) Excluding cross-layer GCL would degrade the performance significantly. This verifies that the proposed cross-layer GCL is indeed effective in enhancing intra-class compactness and inter-class separability. 6) Removing OOD score regularization weakens the performance, reflecting its effectiveness on improving the discriminability of OOD scores between ID and OOD nodes. 7) Without clustering-then-ranking, i.e., directly ranking OOD score instead, would significantly degrade the performance. This verifies that clustering-then-ranking indeed selects more reliable potential-OOD nodes.

\textbf{Visualization of Embeddings}. We adopt t-SNE \cite{tsne} to visualize the embeddings learned by different methods. As shown in Figure \ref{fig.tsne}, compared to competitive baselines GRASP and NODESAFE++, our negMIX forms clearer boundaries between different ID classes, while pushing the OOD class (grey color) farther away from ID classes. This again verifies that negMIX can enhance intra-class compactness and inter-class separability. 

\textbf{Visualization of Conventional Positive Mixup vs. Our Negative Mixup}. As shown in Figure \ref{subfig_conmix}, the positively mixed-up pseudo-OOD training samples constructed by conventional positive Mixup would be dispersed across both ID and OOD regions, unavoidably blurring the ID/OOD boundary. Consistent with Theorem 2, Figure \ref{subfig_negmix} demonstrates that the pseudo-OOD training samples constructed by our negative Mixup are predominantly located within the OOD region while far away from the ID regions, which provides more reliable supervision toward the OOD class and sharpens the ID/OOD boundary.

\textbf{Visualization of OOD Score}. As shown in Figure \ref{fig.Visual}, ID nodes tend to exhibit smaller OOD score, while OOD nodes tend to have larger OOD score, and there is little overlap between the scores of ID and OOD. This demonstrates that the proposed neighborhood-aggregated OOD score scheme upon entropy and confidence can effectively distinguish between OOD and ID.%

\begin{table}
\caption{The ablation study results of negMIX.}
\label{tab:ablation_study}
\centering
\setlength{\tabcolsep}{1.0mm}{
\resizebox{1.0\linewidth}{!}{
    \begin{tabular}{ccc} 
        \hline
         \multirow{2}{*}{\textbf{Model Variants}}&\textbf{AmazonPhoto}&\textbf{WikiCS}\\ 
        \cmidrule{2-3}
         &\multicolumn{2}{c}{\textbf{Accuracy$\uparrow$  /Macro-F1$\uparrow$/AUROC$\uparrow$ /FPR@95$\downarrow$}}         \\
        \hline
        negMIX
        &\textbf{94.5}/\textbf{92.5}/\textbf{99.1}/\textbf{2.0}&\textbf{82.2}/\textbf{81.2}/\textbf{93.3}/\textbf{35.1}\\ 
        
        w/o Positive Mixup for ID
        &93.7/91.8/98.7/6.9&81.5/78.0/92.3/37.2\\ 
        
        Positive Mixup for OOD
        &90.3/89.8/97.0/18.2&81.5/80.9/92.2/38.7\\
        
        Selected OOD w/o Negative Mixup
        &93.4/91.6/98.4/11.9&81.9/80.0/92.6/40.8\\ 
        
        w/o Positive Learning Loss	
        &88.2/88.5/94.2/33.9&81.6/80.1/92.3/42.9\\ 
        
        w/o Negative Learning Loss
        &94.1/92.2/98.8/6.3&81.8/80.2/93.2/36.0\\
 
        w/o Cross-layer GCL
        &86.3/84.9/96.3/34.1&81.3/79.6/92.5/41.0\\ 
        
        w/o OOD Regularization
        &94.0/91.9/99.0/2.2&71.4/75.2/87.5/50.2\\

        w/o Clustering-then-ranking
        &82.6/85.1/90.4/62.1&67.9/70.1/82.6/64.8\\
        \hline
    \end{tabular}
    }}
\end{table}

\section{Conclusion}
We propose a novel method, named negMIX to address OSNC problem. To select highly potential OOD nodes, we devise a clustering-then-ranking strategy upon the neighborhood-aggregated OOD scores, which favors the nodes that are both with high OOD score and representative of the OOD distribution. To enhance the model's generalization to OOD nodes, we propose a novel negative Mixup method specifically crafted for the open-set scenario with theoretical justification. To promote intra-class compactness and inter-class separability, a novel cross-layer GCL module is devised to maximize prototypical mutual information across different distance neighborhoods. Extensive experiments demonstrate significant outperformance of the proposed negMIX over SOTA OSNC methods under different training ratios and various numbers of OOD classes.

\begin{acks}
This work was supported in part by Hainan Provincial Natural Science Foundation of China (No. 322RC570), National Natural Science Foundation of China (Nos. 62362020, 62322203 and 62172052), and the Specific Research Fund of The Innovation Platform for Academicians of Hainan Province (No. YSPTZX202410).
\end{acks}

\clearpage

\bibliographystyle{ACM-Reference-Format}
\balance
\bibliography{myref}

\clearpage

\appendix

\section*{Appendix}

\section{Parameter Sensitivity}
\label{app:sensitivity}
\begin{figure}[h]
\centering
\includegraphics[width=0.95\linewidth]{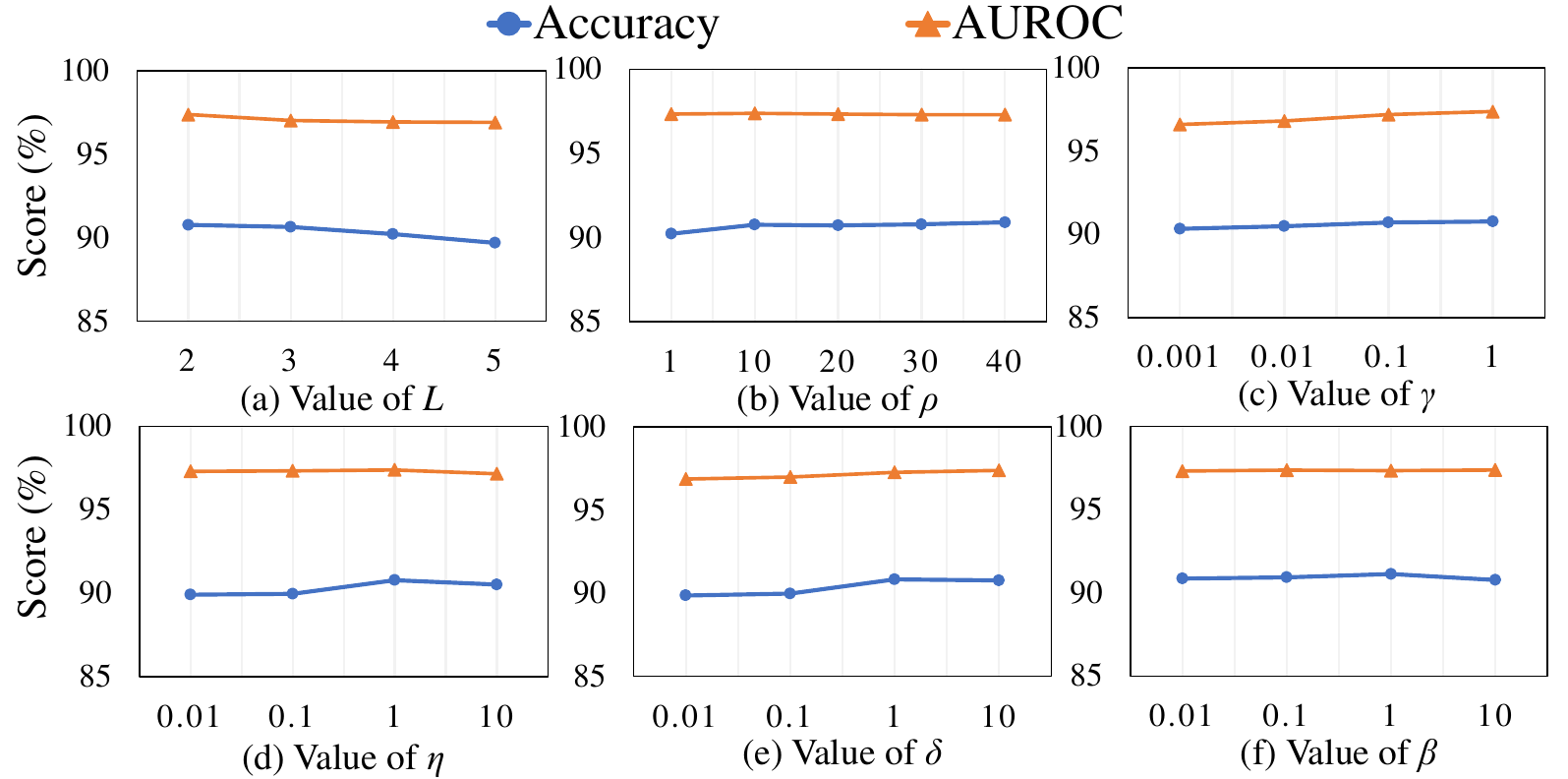}
\caption{Parameter sensitivity on AmazonComputers.} 
\Description{The figure illustrates the parameter sensitivity of negMIX on AmazonComputers.}
\label{fig.parameters}
\end{figure}

As shown in Figure \ref{fig.parameters}, negMIX achieves optimal performance when the number of GNN layers $L$ is set to 2, while deeper GNN encoders would slightly degrade the performance. negMIX is insensitive to $\rho$, i.e., the selection ratio of potential OOD (ID) for Mixup, validating the robustness of the proposed neighborhood-aggregated OOD score scheme and the clustering-then-ranking selection strategy. negMIX is insensitive to the weights of different losses, i.e., OOD score regularization $\gamma$, Mixup loss for training mixed-up pseudo-ID $\eta$, positive and negative learning loss for training mixed-up pseudo-OOD $\delta$, and GCL loss $\beta$.%

\section{Algorithm of negMIX}
\label{app:algorithm}
Algorithm 1 summarizes the training procedure of negMIX. Node embeddings are learned by the GAT encoder and concatenated across layers to form final representations, which are then fed into a $(C{+}1)$-class open-set classifier (lines 2–3).  Cross-entropy loss is applied on labeled nodes (line 4). Clustering-then-ranking is performed on the neighborhood-aggregated OOD scores to select reliable potential ID and OOD nodes for Mixup (lines 5–6). OOD score regularization is imposed to enhance separability between ID and OOD (line 7). The Mixup loss for training mixed-up pseudo-ID nodes and the positive and negative learning loss for training mixed-up pseudo-OOD nodes are jointly optimized to enhance the model's generalization to unseen ID and OOD nodes (lines 8–9). The prototype-to-prototype GCL loss $\mathcal{L}_{p2p}$ is optimized to pull the same class prototypes closer while pushing different class prototypes far apart across GNN layers (line 10). Node-to-prototype GCL loss $\mathcal{L}_{n2p}$ is optimized to explicitly pull nodes toward their corresponding class prototypes while pushing them away from other class prototypes (line 11). The optimized parameters are leveraged for node embedding generation and label prediction after training convergence (lines 14–15).

\begin{figure}[hb]
\flushright
\centering
\includegraphics[width=1\linewidth]{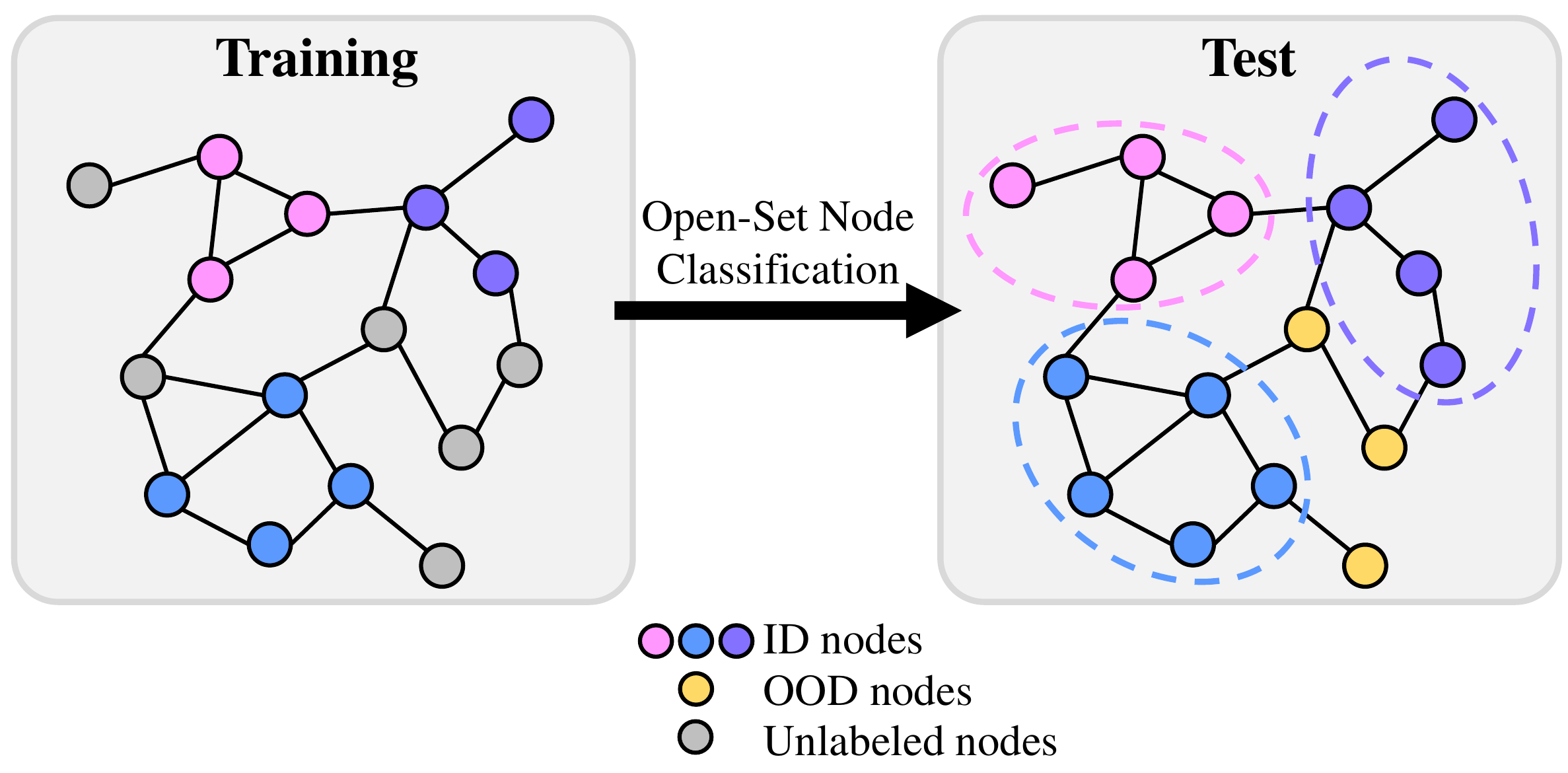}
\caption{An illustration of the open-set node classification problem, where unlabeled nodes may belong to novel classes unseen in the labeled training data. The goal is to accurately classify unlabeled ID nodes into corresponding known classes and reject unlabeled OOD nodes as “unknown” class.}
\Description{An illustration of the open-set node classification problem.}
\label{fig.problem}
\end{figure}

\begin{algorithm}[hb]
    \renewcommand{\algorithmicrequire}{\textbf{Input:}}
	\renewcommand{\algorithmicensure}{\textbf{Output:}}
    \caption{negMIX}
    \label{alg:algorithm}
    \begin{algorithmic}[1] 
        \REQUIRE The whole graph $(\mathcal{V}_{\mathbb{L}},\mathcal{V}_{\mathbb{U}},\boldsymbol{X},\boldsymbol{A},\boldsymbol{Y}_{\mathbb{L}})$ including labeled node set $\mathcal{V}_\mathbb{L}$ and unlabeled node set $\mathcal{V}_\mathbb{U}$.    
        \ENSURE $\text{ Predicted node labels }{\{\hat{y}_i\}_{i=1}^{\mathcal{V}_\mathbb{U}}}.$%
        \WHILE{not max epoch}
        \STATE Learn node embeddings $\{\boldsymbol{h}_i^{(1)},\boldsymbol{h}_i^{(2)},...,\boldsymbol{h}_i^{(L)}\}_{i=1}^N$ and concatenate embeddings of all layers to generate final embeddings.%
        \STATE Feed integrated embeddings to the ($C$+1)-class open-set classifier to generate predicted probabilities.
        \STATE Compute cross-entropy loss $\mathcal{L}_{ce}$ based on $\boldsymbol{Y}_{\mathbb{L}}$.
        \STATE Compute neighborhood-aggregated OOD score in Eq. \ref{eq.ood_score}.
        \STATE Clustering-then-ranking on OOD scores to select a set of potential OOD nodes $\mathcal{V}_{\mathbb{PO}}$ and a set of potential ID nodes $\mathcal{V}_{\mathbb{PI}}$.
        \STATE Compute OOD score regularization $\mathcal{L}_{oreg}$ in Eq. \ref{eq.loss_os}.
        \STATE Conduct Positive Mixup to construct mixed-up pseudo-ID sample in Eq. \ref{eq.mixup_pi} and compute Mixup loss $\mathcal{L}_{pi}$ in Eq. \ref{eq.l_pi}.
        \STATE Conduct negative Mixup to construct mixed-up pseudo-OOD sample in Eq. \ref{eq.mixup_po}, and compute positive and negative learning loss $\mathcal{L}_{po}$ in Eq. \ref{eq.loss_po_p} and \ref{eq.loss_po_n}.
        \STATE Compute prototype-to-prototype contrastive loss $\mathcal{L}_{p2p}$ in Eq. \ref{eq.loss_p2p1}.
        \STATE Compute node-to-prototype contrastive loss $\mathcal{L}_{n2p}$ in Eq. \ref{eq.loss_n2p}.
        \STATE Update learnable parameters to minimize Eq. \ref{eq.total_loss}.
        \ENDWHILE
       \STATE Apply optimized learnable parameters of GNN encoder to generate embeddings.
        \STATE Apply optimized learnable parameters of open-set classifier to generate predicted node labels ${\{\hat{y}_i\}_{i=1}^{\mathcal{V}_\mathbb{U}}}$ in Eq. \ref{eq.hat_y}.
    \end{algorithmic}
\end{algorithm}

\section{Details of Baselines}
\label{app:baselines details}
The proposed negMIX is competed against eight SOTA baselines:
\begin{enumerate}

\item[1)]\textbf{OpenWGL} \cite{OpenWGL} defines a confidence score and an adaptive threshold to detect unseen nodes.

\item[2)]\textbf{OODGAT} \cite{OODGAT} proposes an attention mechanism that facilitates message passing within ID and OOD nodes, while preventing information flow between ID and OOD nodes.

\item[3)]\textbf{GNNSAFE++} \cite{gnnsafe} proposes energy belief propagation to enlarge the margin of energy scores between ID and OOD. 

\item[4)]\textbf{G$^2$Pxy} \cite{G2Pxy} generates unknown class proxies and uses an additional proxy classifier to identify OOD nodes.

\item[5)]\textbf{EMP} \cite{EMP} designs an entropy propagation mechanism to detect unseen-class nodes and transforms threshold selection into a clustering problem.

\item[6)]\textbf{GRASP} \cite{grasp} proposes OOD score propagation, where OOD scores can be defined by both energy and confidence.

\item[7)]\textbf{NODESAFE++} \cite{nodesafe} improves on GNNSAFE++ by bounding negative energy scores. 

\item[8)]\textbf{GOLD} \cite{GOLD} proposes an implicit adversarial learning strategy to synthesize pseudo-OOD samples.
\end{enumerate}

For baselines employing a closed-set classifier without threshold selection, we follow \cite{OODGAT} to report their best scores at the threshold that maximizes Macro-F1. All baselines are implemented following the original papers, using the same hyperparameter settings unless otherwise specified.

\begin{table}[h]
    \caption{Hyper-parameter settings of negMIX on eight datasets. Ratio of selective potential OOD (ID): $\rho$\%; \# heads of each GAT layer: $K$; \# GAT layers: $L$; \# embedding dimensions of each head: $F'$; weight decay: $wd$; learning rate: $lr$; temperature parameter: $\tau$; weight of OOD score regularization: $\gamma$; weight of Mixup loss: $\eta$; weight of positive and negative learning loss: $\delta$; weight of GCL loss: $\beta$.}
    \label{tab:hyperparameter}
\centering
\setlength{\tabcolsep}{0.6mm}{
\resizebox{0.9\linewidth}{!}{
\begin{tabular}{cccccccccccc}
    \toprule
    Dataset&$\rho$&$K$&$L$&$F'$&$wd$&$lr$&$\tau$   &$\gamma$
&$\eta$
&$\delta$&$\beta$\\
    \midrule
Cora
& 10&  2&2&16&1e-3&1e-2&1   &0.1&0.1&1& 1
\\
Citeseer
& 10&  4&2&16&1e-3&1e-2&1   &1&0.1&1& 10
\\
 PubMed
& 10&  4&2& 16& 1e-3& 1e-2& 1   &0.1&0.1& 10& 10
\\
AmazonComputers
& 10&  2&2&16&1e-4&1e-2&1   &1&1&10&10
\\
AmazonPhoto
& 10&  2&2&16&1e-3&1e-2&1   &0.1&0.1&10& 1
\\
 CoauthorCS
& 10&  4&2& 16& 1e-3& 1e-2& 1   &1&0.1& 10& 10
\\
 WikiCS
& 10&  2&2& 16& 1e-3& 1e-2& 1   &1&0.1& 1& 1
\\
 Arxiv
& 10&  4&2& 16& 1e-4& 1e-2& 1   &1&1& 1& 0.1
\\
    \bottomrule
    \end{tabular}}}
\end{table}%

\begin{table}[h]
    \centering
    \caption{Statistics of the datasets.}
    \label{tab:dataset}
    \resizebox{\linewidth}{!}{
    \begin{tabular}{ccccc}
        \toprule
        Dataset&\#Nodes&\#Edges&\#Features&\#Classes\\
        \midrule
        Cora    & 2,708&  10,556&1,433&7    \\
        Citeseer    & 3,327&  9,104&3,703&6    \\
        PubMed    & 19,717&  88,648&500&3    \\
        AmazonComputers    & 13,752&  491,722&767& 10    \\
        AmazonPhoto    & 7,650&  238,162&745& 8    \\
        CoauthorCS    & 18,333&  163,788&6,805&15    \\
        WikiCS    & 11,701&  431,726&300& 10    \\
        Arxiv    & 169,343& 1,166,243& 128& 40    \\
        \bottomrule
    \end{tabular}}
\end{table}

\section{Efficiency Analysis} 
\label{app:efficiency}
We conducted experiments on a hardware platform equipped with an NVIDIA RTX 4060Ti GPU and an Intel(R) Core(TM) i7-13700F CPU. As reported in Table \ref{tab:runningtime}, negMIX completes one training epoch in only 0.3742 seconds on the large-scale benchmark Arxiv, with a peak GPU memory usage of 2521.10 MB, demonstrating both high computational efficiency and moderate memory footprint. Compared to baselines, negMIX achieves a favorable balance between scalability and effectiveness.

\begin{table}[H]
\caption{Training time and Memory on large-scale Arxiv.}
\label{tab:runningtime}
\centering
\setlength{\tabcolsep}{0.8mm}{
\resizebox{0.90\linewidth}{!}{
    \begin{tabular}{ccc} 
        \hline
        \textbf{Method}&\textbf{Per-epoch (s)} &\textbf{Peak GPU memory (MB)}\\ 
        \hline
        OpenWGL &OOM  &OOM\\
        
        OODGAT &0.1142 &4457.35\\ 
        
        G$^2$Pxy &75.373 &6883.71\\ 
        
        GNNSAFE++ &2.7816 &1115.56\\ 
        
        EMP &OOM &OOM\\ 

        GRASP &0.0719 &1520.01\\ 
        
        NODESAFE++ &5.7909 &1190.26\\ 

        GOLD &OOM &OOM\\ 

        negMIX (Ours) &0.3742 &2521.10\\
        \hline
    \end{tabular}}
    }
\end{table}





\section{Proof of Theorem 1}
\label{app:proof1}

\subsection{Preliminary}

\begin{flalign}
\displaystyle
&\boldsymbol{h}_i^{\mathbb{MO}}=\lambda\boldsymbol{h}_i^{\mathbb{PO}}+(1-\lambda)\boldsymbol{h}_j^{\mathbb{L}},\quad \boldsymbol{y}_i^{\mathbb{MO}}=\lambda\bar{\boldsymbol{y}}_i^{\mathbb{PO}}+(1-\lambda)\boldsymbol{y}_j^{\mathbb{L}}\notag&
 \end{flalign}%
\begin{flalign}
    \displaystyle
&\mathcal{L}_{po}=-\frac{1}{|\mathcal{V}_{\mathbb{PO}}|}
\sum\nolimits_{i=1}^{|\mathcal{V}_{\mathbb{PO}}|}\left(\lambda\log{p}_{i,C+1}^{\mathbb{MO}}+(1-\lambda)\log{p}_{i,y_j}^{\mathbb{MO}}\right)\notag&
\end{flalign}%
For ease of subsequent notations, we denote $\boldsymbol{h}_{i}^{\mathbb{PO}}$ as $\boldsymbol{h}_{i}$, ${p}_{i,k}^{\mathbb{PO}}$ as ${p}_{i,k}$, $\boldsymbol{h}_{j}^{\mathbb{L}}$ as $\boldsymbol{h}_{j}$, ${p}_{j,k}^{\mathbb{L}}$ as ${p}_{j,k}$, $\boldsymbol{h}_{i}^{\mathbb{MO}}$ as $\check{\boldsymbol{h}}_{i}$, ${p}_{i,k}^{\mathbb{MO}}$ as $\check{p}_{i,k}$. %
For any $\boldsymbol{s}_i$, we have:
\begin{flalign}
&\frac{\partial\boldsymbol{s}_i}{\partial\boldsymbol{h}_i}\frac{\partial\boldsymbol{h}_i}{\partial\boldsymbol{s}_i}=\boldsymbol{I},\quad \check{\boldsymbol{s}}_i=\boldsymbol{W}\cdot\check{\boldsymbol{h}}_i,\quad \check{p}_{i,k}={\exp(\check{s}_{i,k})}/{\sum\nolimits_{c=1}^{C+1}\exp(\check{s}_{i,c})}\notag&
\end{flalign}

\begin{flalign}
&\frac{\partial\check{\boldsymbol{s}}_i}{\partial\check{\boldsymbol{h}}_i}\cdot\frac{\partial\check{\boldsymbol{h}}_i}{\partial\boldsymbol{h}_j}\cdot\frac{\partial\boldsymbol{h}_j}{\partial\boldsymbol{s}_j}=(1-\lambda)\boldsymbol{I},\quad\frac{\partial\check{\boldsymbol{s}}_i}{\partial\check{\boldsymbol{h}}_i}\cdot\frac{\partial\check{\boldsymbol{h}}_i}{\partial\boldsymbol{h}_i}\cdot\frac{\partial\boldsymbol{h}_i}{\partial\boldsymbol{s}_i}=\lambda\boldsymbol{I}\notag&
\end{flalign}
\begin{flalign}
&\frac{\partial \check{p}_{i,C+1}}{\partial \check{\boldsymbol{s}}_i} = (-\check{p}_{i,C+1} \check{p}_{i,1}, ..., \check{p}_{i,C+1} (1 - \check{p}_{i,C+1}))\notag\\
&\frac{\partial \check{p}_{i,y_j}}{\partial \check{\boldsymbol{s}}_i} = 
(-\check{p}_{i,y_j} \check{p}_{i,1}, ..., \check{p}_{i,y_j} (1 - \check{p}_{i,y_j}), ..., -\check{p}_{i,y_j} \check{p}_{i,C+1})\notag\\
&\frac{\partial \boldsymbol{s}_i}{\partial p_{i,y_j}} = (\frac{-1}{p_{i,y_j} p_{i,1}}, ..., \frac{1}{p_{i,y_j} (1 - p_{i,y_j})},..., \frac{-1}{p_{i,y_j} p_{i,C+1}})^\mathrm{T}\notag\\
&\frac{\partial \boldsymbol{s}_i}{\partial p_{i,C+1}} = (\frac{-1}{p_{i,C+1} p_{i,1}},\frac{-1}{p_{i,C+1} p_{i,2}}, ..., \frac{1}{p_{i,C+1}(1-p_{i,C+1})})^\mathrm{T}\notag&
\end{flalign}%

\subsection{Analyze the Influence on the Predictions}

(1) Minimizing  ($-\lambda\log\check{p}_{i,C+1}$) increases unknown class probability $p_{i,C+1}$ for potential OOD $v_i$:

\begin{flalign}
\displaystyle
&\frac{\partial(-\lambda\mathrm{log}\check{p}_{i,C+1})}{\partial p_{i,C+1}}\notag\\
&=\frac{\partial(-\lambda\mathrm{log}\check{p}_{i,C+1})}{\partial\check{p}_{i,C+1}}\cdot\frac{\partial\check{p}_{i,C+1}}{\partial\check{\boldsymbol{s}}_i}\cdot\frac{\partial\check{\boldsymbol{s}}_i}{\partial\check{\boldsymbol{h}}_i}\cdot\frac{\partial\check{\boldsymbol{h}}_i}{\partial\boldsymbol{h}_i}\cdot\frac{\partial\boldsymbol{h}_i}{\partial\boldsymbol{s}_i}\cdot\frac{\partial\boldsymbol{s}_i}{\partial p_{i,C+1}}\notag\\
&=\frac{-\lambda\cdot\lambda}{\check{p}_{i,C+1}}\cdot(-\check{p}_{i,C+1}\check{p}_{i,1},...,\check{p}_{i,C+1}(1-\check{p}_{i,C+1}))\notag\\
&\cdot(\frac{-1}{p_{i,C+1}p_{i,1}},\frac{-1}{p_{i,C+1}p_{i,2}},...,\frac{1}{p_{i,C+1}(1-p_{i,C+1})})^\mathrm{T}\notag\\
&=\frac{-\lambda^2}{{p}_{i,C+1}}(\frac{\check{p}_{i,1}}{p_{i,1}}+\frac{\check{p}_{i,2}}{p_{i,2}}+...+\frac{1-\check{p}_{i,1}}{(1-p_{i,C+1})})<0\notag&
\end{flalign}%

(2) Minimizing ($-\lambda\log\check{p}_{i,C+1}$) increases unknown class probability $p_{j,C+1}$ for labeled-ID $v_j$:%

\begin{flalign}
\displaystyle
&\frac{\partial(-\lambda\mathrm{log}\check{p}_{i,C+1})}{\partial p_{j,C+1}}\notag\\
&=\frac{\partial(-\lambda\mathrm{log}\check{p}_{i,C+1})}{\partial\check{p}_{i,C+1}}\cdot\frac{\partial\check{p}_{i,C+1}}{\partial\check{\boldsymbol{s}}_i}\cdot\frac{\partial\check{\boldsymbol{s}}_i}{\partial\check{\boldsymbol{h}}_i}\cdot\frac{\partial\check{\boldsymbol{h}}_i}{\partial\boldsymbol{h}_j}\cdot\frac{\partial\boldsymbol{h}_j}{\partial\boldsymbol{s}_j}\cdot\frac{\partial\boldsymbol{s}_j}{\partial p_{j,C+1}}\notag\\
&=\frac{-\lambda\cdot(1-\lambda)}{\check{p}_{i,C+1}}\cdot(-\check{p}_{i,C+1}\check{p}_{i,1},...,\check{p}_{i,C+1}(1-\check{p}_{i,C+1}))\notag\\
&\cdot(\frac{-1}{p_{j,C+1}p_{j,1}},\frac{-1}{p_{j,C+1}p_{j,2}},...,\frac{1}{p_{j,C+1}(1-p_{j,C+1})})^\mathrm{T}\notag\\
&=\frac{\lambda\cdot(\lambda-1)}{{p}_{j,C+1}}(\frac{\check{p}_{i,1}}{p_{j,1}}+\frac{\check{p}_{i,2}}{p_{j,2}}+...+\frac{1-\check{p}_{i,C+1}}{(1-p_{j,C+1})})<0\notag&
\end{flalign}%

(3) Minimizing $(\lambda-1)\log\check{p}_{i,y_j}$ increases known class probability $p_{j,y_j}$ for labeled-ID $v_j$:

\begin{flalign}
\displaystyle
&\frac{\partial((\lambda-1)\mathrm{log}\check{p}_{i,y_j})}{\partial p_{j,y_j}}\notag\\
&=\frac{\partial((\lambda-1)\mathrm{log}\check{p}_{i,y_j})}{\partial\check{p}_{i,y_j}}\cdot\frac{\partial\check{p}_{i,y_j}}{\partial\check{\boldsymbol{s}}_i}\cdot\frac{\partial\check{\boldsymbol{s}}_i}{\partial\check{\boldsymbol{h}}_i}\cdot\frac{\partial\check{\boldsymbol{h}}_i}{\partial\boldsymbol{h}_j}\cdot\frac{\partial\boldsymbol{h}_j}{\partial\boldsymbol{s}_j}\cdot\frac{\partial\boldsymbol{s}_j}{\partial p_{j,y_j}}\notag\\
&=\frac{-(1-\lambda)^2}{\check{p}_{i,y_j}}\cdot(-\check{p}_{i,y_j}\check{p}_{i,1},...,\check{p}_{i,y_j}(1-\check{p}_{i,y_j}),...,-\check{p}_{i,y_j}\check{p}_{i,C+1})\notag\\
&\cdot(\frac{-1}{p_{j,y_j}p_{j,1}},...,\frac{1}{p_{j,y_j}(1-p_{j,y_j})},...,\frac{-1}{p_{j,y_j}p_{j,C+1}})^\mathrm{T}\notag\\
&=\frac{-(1-\lambda)^2}{p_{j,y_j}}(\frac{\check{p}_{i,1}}{p_{j,1}}+...+\frac{(1-\check{p}_{i,y_j})}{(1-p_{j,y_j})}+...+\frac{\check{p}_{i,C+1}}{p_{j,C+1}})<0\notag&
\end{flalign}%

(4) Minimizing $(\lambda-1)\log\check{p}_{i,y_j}$ increases known class probability $p_{i,y_j}$ for potential OOD $v_i$:%
\begin{flalign}
\displaystyle
&\frac{\partial((\lambda-1)\log\check{p}_{i,y_j})}{\partial p_{i,y_j}}\notag\\
&=\frac{\partial((\lambda-1)\log\check{p}_{i,y_j})}{\partial\check{p}_{i,y_j}}\cdot\frac{\partial\check{p}_{i,y_j}}{\partial\check{\boldsymbol{s}}_i}\cdot\frac{\partial\check{\boldsymbol{s}}_i}{\partial\check{\boldsymbol{h}}_i}\cdot\frac{\partial\check{\boldsymbol{h}}_i}{\partial\boldsymbol{h}_i}\cdot\frac{\partial\boldsymbol{h}_i}{\partial\boldsymbol{s}_i}\cdot\frac{\partial\boldsymbol{s}_i}{\partial p_{i,y_j}}\notag\\
&=\frac{\lambda(\lambda-1)}{\check{p}_{i,y_j}}\cdot(-\check{p}_{i,y_j}\check{p}_{i,1},...,\check{p}_{i,y_j}(1-\check{p}_{i,y_j})...,-\check{p}_{i,y_j}\check{p}_{i,C+1})\notag\\
&\cdot(\frac{-1}{p_{i,y_j}p_{i,1}},...,\frac{1}{p_{i,y_j}(1-p_{i,y_j})},...,\frac{-1}{p_{i,y_j}p_{i,C+1}})^\mathrm{T}\notag\\
&=\frac{\lambda(\lambda-1)}{p_{i,y_j}}(\frac{\check{p}_{i,1}}{p_{i,1}}+...+\frac{(1-\check{p}_{i,y_j})}{(1-p_{i,y_j})}+...+\frac{\check{p}_{i,C+1}}{p_{i,C+1}})<0\notag&
\end{flalign}%

The above inference reveals that conventional Mixup jointly increases both known and unknown class probabilities for potential OOD and labeled ID samples. Specifically, while minimizing $-\lambda\log\check{p}_{i,C+1}$ raises the unknown-class probability $p_{i,C+1}$ for potential OOD sample $v_i$, it undesirably increases $p_{j,C+1}$ for labeled-ID sample $v_j$. Minimizing $-(1-\lambda)\log\check{p}_{i,y_j}$ strengthens the known-class probability $p_{j,y_j}$ for labeled-ID samples as intended, but also elevates $p_{i,y_j}$ for potential OOD samples. Consequently, the ID/OOD decision boundary becomes blurred.

\section{Proof of Theorem 2}
\label{app:proof2}
\subsection{Preliminary}

\begin{flalign}
\displaystyle
&\boldsymbol{h}_i^{\mathbb{MO}}=\lambda\boldsymbol{h}_i^{\mathbb{PO}}+(1-\lambda)(-\boldsymbol{h}_j^{\mathbb{L}}), \quad \boldsymbol{y}_i^{\mathbb{MO}}=\lambda\bar{\boldsymbol{y}}_i^{\mathbb{PO}}+(1-\lambda)(-\boldsymbol{y}_j^{\mathbb{L}})\notag&
 \end{flalign}%
\begin{flalign}
\displaystyle
&\mathcal{L}_{po}=\frac{1}{|\mathcal{V}_{\mathbb{PO}}|}
\sum\nolimits_{i=1}^{|\mathcal{V}_{\mathbb{PO}}|}(-\lambda\log{p}_{i,C+1}^{\mathbb{MO}}-(1-\lambda)\log(1-{p}_{i,y_j}^{\mathbb{MO}}))\notag&
\end{flalign}%
 \begin{flalign}
&\frac{\partial\check{\boldsymbol{s}}_i}{\partial\check{\boldsymbol{h}}_i}\cdot\frac{\partial\check{\boldsymbol{h}}_i}{\partial\boldsymbol{h}_j}\cdot\frac{\partial\boldsymbol{h}_j}{\partial\boldsymbol{s}_j}=-(1-\lambda)\boldsymbol{I};  \frac{\partial\check{\boldsymbol{s}}_i}{\partial\check{\boldsymbol{h}}_i}\cdot\frac{\partial\check{\boldsymbol{h}}_i}{\partial\boldsymbol{h}_i}\cdot\frac{\partial\boldsymbol{h}_i}{\partial\boldsymbol{s}_i}=\lambda\boldsymbol{I}\notag&
\end{flalign}

\subsection{Analyze the Influence on the Predictions}

(1) Minimizing positive learning loss ($-\lambda\log\check{p}_{i,C+1}$) increases unknown class probability $p_{i,C+1}$ for potential OOD $v_i$:
\begin{flalign}
\displaystyle
&\frac{\partial(-\lambda\mathrm{log}\check{p}_{i,C+1})}{\partial p_{i,C+1}}\notag\\
&=\frac{\partial(-\lambda\mathrm{log}\check{p}_{i,C+1})}{\partial\check{p}_{i,C+1}}\cdot\frac{\partial\check{p}_{i,C+1}}{\partial\check{\boldsymbol{s}}_i}\cdot\frac{\partial\check{\boldsymbol{s}}_i}{\partial\check{\boldsymbol{h}}_i}\cdot\frac{\partial\check{\boldsymbol{h}}_i}{\partial\boldsymbol{h}_i}\cdot\frac{\partial\boldsymbol{h}_i}{\partial\boldsymbol{s}_i}\cdot\frac{\partial\boldsymbol{s}_i}{\partial p_{i,C+1}}\notag\\
&=\frac{-\lambda^2}{{p}_{i,C+1}}(\frac{\check{p}_{i,1}}{p_{i,1}}+\frac{\check{p}_{i,2}}{p_{i,2}}+...+\frac{1-\check{p}_{i,1}}{(1-p_{i,C+1})})<0\notag&
\end{flalign}%

(2) Minimizing positive learning loss ($-\lambda\log\check{p}_{i,C+1}$) decreases unknown class probability $p_{j,C+1}$ for labeled-ID $v_j$:
\begin{flalign}
&\frac{\partial(-\lambda\mathrm{log}\check{p}_{i,C+1})}{\partial p_{j,C+1}}\notag\\
&=\frac{\partial(-\lambda\mathrm{log}\check{p}_{i,C+1})}{\partial\check{p}_{i,C+1}}\cdot\frac{\partial\check{p}_{i,C+1}}{\partial\check{\boldsymbol{s}}_i}\cdot\frac{\partial\check{\boldsymbol{s}}_i}{\partial\check{\boldsymbol{h}}_i}\cdot\frac{\partial\check{\boldsymbol{h}}_i}{\partial\boldsymbol{h}_j}\cdot\frac{\partial\boldsymbol{h}_j}{\partial\boldsymbol{s}_j}\cdot\frac{\partial\boldsymbol{s}_j}{\partial p_{j,C+1}}\notag\\
&=\frac{\lambda\cdot(1-\lambda)}{{p}_{j,C+1}}(\frac{\check{p}_{i,1}}{p_{j,1}}+\frac{\check{p}_{i,2}}{p_{j,2}}+...+\frac{1-\check{p}_{i,C+1}}{(1-p_{j,C+1})})>0\notag&
\end{flalign}%

(3) Minimizing negative learning loss $(\lambda-1)\log(1-\check{p}_{i,y_j})$ increases known class probability $p_{j,y_j}$ for labeled-ID $v_j$:%
\begin{flalign}
\displaystyle
&\frac{\partial((\lambda-1)\mathrm{log}(1-\check{p}_{i,y_j}))}{\partial p_{j,y_j}}\notag\\
&=\frac{\partial((\lambda-1)\mathrm{log}(1-\check{p}_{i,y_j}))}{\partial\check{p}_{i,y_j}}\cdot\frac{\partial\check{p}_{i,y_j}}{\partial\check{\boldsymbol{s}}_i}\cdot\frac{\partial\check{\boldsymbol{s}}_i}{\partial\check{\boldsymbol{h}}_i}\cdot\frac{\partial\check{\boldsymbol{h}}_i}{\partial\boldsymbol{h}_j}\cdot\frac{\partial\boldsymbol{h}_j}{\partial\boldsymbol{s}_j}\cdot\frac{\partial\boldsymbol{s}_j}{\partial p_{j,y_j}}\notag\\
&=\frac{-(1-\lambda)^2\check{p}_{i,y_j}}{(1-\check{p}_{i,y_j})p_{j,y_j}}(\frac{\check{p}_{i,1}}{p_{j,1}}+...+\frac{(1-\check{p}_{i,y_j})}{(1-p_{j,y_j})}+...+\frac{\check{p}_{i,C+1}}{p_{j,C+1}})<0\notag&
\end{flalign}%

(4) Minimizing negative learning loss $(\lambda-1)\log(1-\check{p}_{i,y_j})$ decreases known class probability $p_{i,y_j}$ for potential OOD $v_i$:%
\begin{flalign}
&\frac{\partial((\lambda-1)\log(1-\check{p}_{i,y_j}))}{\partial p_{i,y_j}}\notag\\
&=\frac{\partial((\lambda-1)\log(1-\check{p}_{i,y_j}))}{\partial\check{p}_{i,y_j}}\cdot\frac{\partial\check{p}_{i,y_j}}{\partial\check{\boldsymbol{s}}_i}\cdot\frac{\partial\check{\boldsymbol{s}}_i}{\partial\check{\boldsymbol{h}}_i}\cdot\frac{\partial\check{\boldsymbol{h}}_i}{\partial\boldsymbol{h}_i}\cdot\frac{\partial\boldsymbol{h}_i}{\partial\boldsymbol{s}_i}\cdot\frac{\partial\boldsymbol{s}_i}{\partial p_{i,y_j}}\notag\\
&=\frac{(1-\lambda)\lambda\check{p}_{i,y_j}}{(1-\check{p}_{i,y_j})p_{i,y_j}}(\frac{\check{p}_{i,1}}{p_{i,1}}+...+\frac{(1-\check{p}_{i,y_j})}{(1-p_{i,y_j})}+...+\frac{\check{p}_{i,C+1}}{p_{i,C+1}})>0\notag&
\end{flalign}%

The above inference steps demonstrate that positive and negative loss design upon negative Mixup promotes mutually exclusive ID/OOD decision behavior, ultimately leading to a clearer boundary between ID and OOD. For potential OOD nodes, it increases their unknown class probability while decreasing known class probability. For labeled ID nodes, it increases their known class probability while decreasing unknown class probability. This dual effect results in a clearer separation between ID and OOD data, ultimately facilitating both accurate ID classification and OOD detection.

\end{document}